\documentclass[useAMS,usenatbib]{mn2e}
\usepackage{graphicx}
\usepackage{epsfig} 
\usepackage{multirow} 
\usepackage{aas_macros}

\title[Star formation in CLG0218 at $z$=1.62 and its large scale environment]
{Star formation in the cluster CLG0218.3-0510 at \textit{z}=1.62 and its large-scale environment: the infrared perspective}
\author[Joana S. Santos et al.]
{J. S. Santos$^{1}$\thanks{E-mail:jsantos@sciops.esa.int}, B. Altieri$^{1}$, M. Tanaka$^{2}$, I. Valtchanov$^{1}$, A. Saintonge$^{3}$,  
M. Dickinson$^{4}$, \and S. Foucaud$^{5}$, T. Kodama$^{2}$, T. D. Rawle$^{1}$, K. Tadaki$^{2}$ \\
$^{1}$European Space Astronomy Centre (ESAC)/ESA, Villanueva de la Ca\~nada, 28691, Madrid, Spain\\
$^{2}$National Astronomical Observatory of Japan, 2-21-1 Osawa, Mitaka, Tokyo 181-8588, JAPAN \\ 
$^{3}$Max-Planck-Institut f\"ur extraterrestrische Physik, Giessenbachstra\ss e, 85748 Garching, Germany\\
$^{4}$National Optical Astronomy Observatory, 950 North Cherry Avenue, Tucson, AZ 85719, USA  \\
$^{5}$Center for Astronomy \& Astrophysics, Department of Physics \& Astronomy, Shanghai Jiao Tong University, 800 Dongchuan Rd., \\ 
Shanghai, 200240, China \\
}

\begin{document}

\date{Accepted . Received; in original form}

\pagerange{\pageref{firstpage}--\pageref{lastpage}} \pubyear{2002}

\maketitle

\label{firstpage}

\begin{abstract}
The galaxy cluster CLG0218.3-0510 at $z$=1.62 is one of the most distant galaxy clusters known, with a rich muti-wavelength
data set that confirms a mature galaxy population already in place.
Using very deep, wide area (20$\times$20 Mpc) imaging by \textit{Spitzer} MIPS at 24$\mu m$, in conjunction with \textit{Herschel} 
5-band imaging from 100--500$\mu m$, we investigate the dust-obscured, star-formation properties in the cluster and its associated 
large scale environment.
 Our galaxy sample of 693 galaxies at $z\sim$1.62 detected at 24$\mu m$ (10 spectroscopic and 683 photo-$z$) includes both cluster galaxies 
(i.e. within $r<$1 Mpc projected clustercentric radius) and field galaxies, defined as the region beyond a radius of 3 Mpc. 
The star-formation rates (SFRs) derived from the measured infrared luminosity range from 18 to 2500 M$_\odot$/yr, with a median of 55 M$_\odot$/yr, 
{over the entire radial range (10 Mpc).
The cluster brightest FIR galaxy, taken as the centre of the galaxy system, is vigorously forming stars at a rate of 
256$\pm$70 M$_\odot$/yr, and the total cluster SFR enclosed in a circle of 1 Mpc is 1161$\pm$96 M$_\odot$/yr. 
We estimate a dust extinction of $\sim$ 3 magnitudes 
by comparing the SFRs derived from [OII] luminosity with the ones computed from the 24$\mu m$ fluxes. 
We find that the in-falling region (1--3 Mpc) is special: there is a significant decrement (3.5$\times$) of passive relative to star-forming 
galaxies in this region, and the total SFR of the galaxies located in this region is lower ($\sim$130 M$_\odot$/yr/Mpc$^{2}$) than 
anywhere in the cluster or field, regardless of their stellar mass.
In a complementary approach we compute the local galaxy density, $\Sigma_{5}$, and find no trend between SFR and $\Sigma_{5}$. 
However, we measure an excess of star-forming galaxies in the cluster relative to the field by a factor 1.7, that 
lends support to a reversal of SF--density relation in CLG0218. }

\end{abstract}

\begin{keywords}
Galaxy clusters - high redshift: observations - FIR: Galaxy clusters - individual - CLG 0218.3-0510 : star formation
\end{keywords}

\section{Introduction}

One of the main concerns of modern astrophysics is the lack of a detailed understanding of galaxy formation in a 
cosmological context. The star-formation rate (SFR) of galaxies in different environments and across cosmic time 
is one of the key physical parameters to understand their evolution.
  
In the local universe, luminous infrared galaxies (LIRGs,  10$^{11} < L_{IR}<$10$^{12}$ L$_\odot$) avoid the high-density environments of 
galaxy clusters in which star formation has already been quenched, and ultra luminous infrared galaxies (ULIRGs, $L_{IR}>$10$^{12}$ $L_\odot$) 
are virtually absent up to $z\sim$0.5 \citep [e.g.,][] {Magnelli, Haines13}. 
Since the peak of the cosmological star-formation rate occurred at  1 $< z <$ 3 \citep{Madau}, it is expected that star 
formation in the denser environments must at some time 
increase towards higher redshifts. However, the amount, spatial distribution, and timing of this increase remains unclear. 

\begin{figure*}
\includegraphics[width=9.7cm,angle=0]{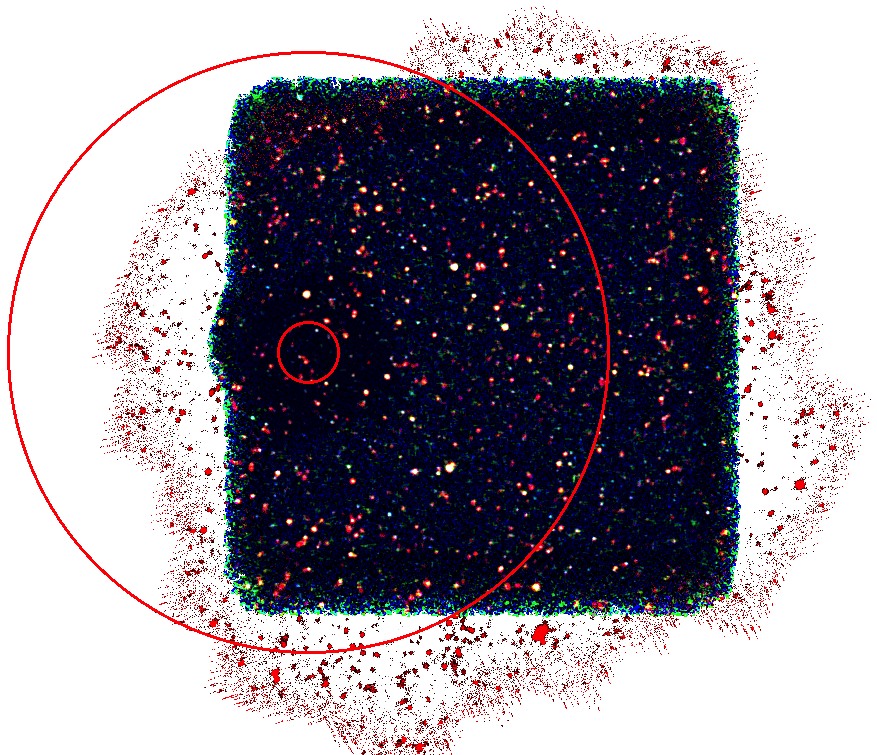} 
\hspace{0.5cm}
\includegraphics[width=7.cm,angle=0]{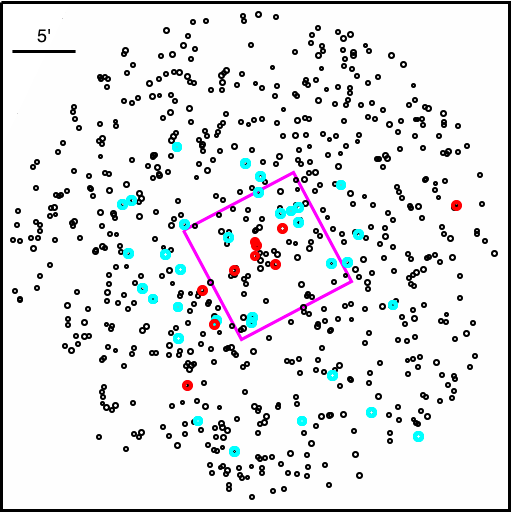} 
 \caption{(\textit{Left}) Colour image (250/160/100 $\mu m$) of the \textit{Herschel} co-added observations of CLG0218. The inner red circle
 is centered on the galaxy ID 16 and has a radius of 2$\arcmin$. The larger circle with $r$=10 Mpc encloses the area covered by the MIPS 
 catalogue used in this work.
 (\textit{Right}) Footprint of cluster members with MIPS detection in a 10 Mpc radius area centered on the cluster galaxy ID 16. 
 Photometric redshift galaxies with $z_{phot}$= 1.52-1.72 are shown in black, spectroscopic galaxies in red and $z_{phot}$ galaxies 
 with [OII] narrow band emission in cyan. The magenta 10$\arcmin \times$10$\arcmin$ square corresponds 
 to the area with deep PACS imaging. 
 }
 \label{herschelimage}
\end{figure*}

The evolution of the star-formation rate with redshift has been studied up to $z\sim$1 and is typically parametrized by a power-law function, 
SFR($z$) $\propto$ (1+$z$)$^{\alpha}$, with a slope $\alpha$ ranging from 2 to 7 \citep[e.g.,][]{Kodama, Geach, Saintonge, Bai, Haines}. 
The recent investigation of \cite{Webb} using 42 clusters from the Red Sequence Cluster Survey lends support to a slope of 5--6 and ascribes this 
evolutionary trend to the in-falling galaxies in clusters at $z>$ 0.75. This conclusion underlines the importance of environment as we probe higher redshifts.
Beyond redshift unity the evolution of the SFR in clusters is unconstrained mostly because of the scarcity of well studied systems 
(i.e., few confirmed clusters with good multi-wavelength data), however the recent works of \citep{Brodwin, Zeimann} made a first step in that regime.

A reversal of the star formation--density relationship at $z >$ 1 was reported by \cite{Elbaz,Elbaz11}  in \textit{low} galaxy density environments (i.e. the field) 
where, unlike in the local Universe, the average star-formation rate of galaxies increases with local galaxy density. 
On the other hand, mid-infrared observations of a couple of galaxy clusters at the highest known redshifts
($z\sim$ 1.5) have revealed a population of IR luminous, actively star-forming galaxies in the cluster cores 
\citep{Hilton, Tran10}, indicating a reversal of the SF--density relation in higher galaxy density systems.

 At intermediate and high-redshifts most of the energy from star formation and AGN activity is absorbed by dust and re-radiated 
at IR wavelengths.  The \textit{Infrared Space Observatory} (ISO) revealed that in intermediate redshift clusters 
most of the star formation is hidden at optical wavelengths \citep [e.g.,][] {Duc, Metcalfe}.  
Subsequent observations with \textit{Spitzer}/MIPS confirmed this scenario \citep [e.g.,][] {Geach, Marcillac}. 
Specifically, \cite{Saintonge} observed an increasing fraction of dusty star-forming cluster galaxies from 3\% locally 
to 13\% at $z$ = 0.83  \citep [see also][] {Haines}. In addition, it has been shown that most of the dust-obscured star formation 
in z $<$ 0.85 clusters happens in intermediate density regions \citep{Koyama} or in groups \citep{Tran09}. 

The \textit{Herschel} space observatory \citep{Pilbratt} is the largest space telescope to date and 
provides unrivalled sensitivity in the wavelength range 55 to 672$\mu m$. Therefore, \textit{Herschel} brackets the 
critical peak of FIR emission of $z\sim$1-2 galaxies, providing a direct, unbiased measurement of star formation. 
Up to now only a handful of studies of high-$z$ clusters and proto-clusters using \textit{Herschel} data have been published 
\citep{Popesso, Seymour, Santos, Pintos-Castro, Alberts}, mostly because the angular resolution at the longest wavelengths 
(up to 18$\arcsec$) does not allow one to resolve small distant galaxies, and furthermore most of the observations have a high SFR detection threshold, 
enabling the study of the highly star-forming galaxies only. 
To investigate the relation between star-formation activity and environment at the epoch when clusters are assembling galaxies 
and galaxies are still undergoing their own formation process, several \textit{Herschel} programmes, namely the Key Project PEP 
\citep[PACS Evolutionary Probe,][] {Lutz}, and several guaranteed time (GT, PI Altieri) and open time (OT) programmes (PI Pope, 
Popesso) have targeted several tens of high-redshift clusters in a broad range of halo masses.

In this paper we present a detailed characterization of the infrared star-formation properties in the galaxy population 
of CLG0218.3-0510 (hereafter CLG0218, RA=34.59955  DEC= -5.17385) at $z$=1.62 and its large scale environment. 
This distant galaxy system was independently discovered as a strong overdensity of red galaxies using \textit{Spitzer}-IRAC 
imaging \citep{Papovich} and as a weak X-ray detection in XMM-Newton data \citep{Tanaka10}. Subsequently, CLG0218 was 
observed with deep, multi-wavelength follow-up observations using the major astronomical observatories, in particular the \textit{Spitzer}
 public legacy survey (SpUDS, PI: Dunlop) and the Cosmic Assembly Near-Infrared Deep 
Extragalactic Legacy Survey (CANDELS, PIs Faber, Ferguson). While the number of confirmed cluster members has been steadily increasing 
with a current sample of $\sim$ 50 galaxies \citep{Papovich, Tanaka10, Tadaki}, the analysis of 85 ksec X-ray \textit{Chandra} 
\citep{Pierre} data rendered inconclusive whether there is extended emission associated with the galaxy overdensity, 
mostly because a bright point source associated with the central cluster galaxy dominates the X-ray flux. Removing this point 
source an upper limit on the cluster mass of 7.7$\times$10$^{13}$M$_\odot$ was set, placing CLG0218 in the 
low mass / group category, or even proto-cluster. This value is consistent with the one derived from the deeper XMM-Newton data 
published in \cite{Tanaka10}, 5.7$\pm$1.4$\times$10$^{13}$M$_\odot$, which we adopt in this paper.
 Throughout this paper we will simply refer to CLG0218 as a cluster of galaxies. 
The star-formation properties of this cluster have been studied using MIPS-24$\mu m$ in a central region of 1 Mpc by \cite{Tran10}.
They found an enhancement of \textit{the fraction} of star-forming galaxies in the cluster relative to lower redshift clusters. One caveat in this 
study is that the 24$\mu m$ derived SFRs were computed with the \cite{Chary} templates which are known to overestimate the SFR for galaxies 
above $z$=1.5. \cite{Tadaki} followed a different approach using near-infrared (NIR) narrow-band imaging targeted to detect [OII] emitters 
in the cluster and the surrounding environment. They found a large filamentary structure around CLG0218 traced by [OII] emitters with a measured 
overdensity a factor 10 larger in the high density regions (cluster core and clumps) than in the field.

Here we use the largest sample of $z_{phot}$ and spectroscopically confirmed galaxies, and infrared maps that cover an area of 20$\times$20 Mpc 
thus enabling the study of the dust-obscured star formation in the cluster and its large scale environment.
For a subset of the galaxy sample we have robust \textit{Herschel} measurements which firmly constrains the amount of dust extinction and validates 
the star formation rates from 24$\mu m$.

The paper is organized as follows: 
in \S 2 we describe the MIPS, PACS and SPIRE observations and reduction procedures while in \S 4 we describe the multi-$\lambda$ ancillary 
used in this work.
In \S 5 we obtain total infrared luminosities ($L_{IR}$) and derive star formation rates (SFRs) from the 24$\mu m$ data. 
A detailed analysis of the infrared properties of the central, IR brightest cluster galaxy is presented in \S 6.
In \S 7 and \S 8 we investigate the relations between the environment, stellar mass and star formation.
Our conclusions are summarized in \S 9.  

The cosmological parameters used throughout the paper are: $H_{0}$=70 $h$ km/s/Mpc ($h$=1),
$\Omega_{\Lambda}$=0.7 and $\Omega_{\rm m}$=0.3. In this cosmology, 1 Mpc at $z$=1.6 corresponds to $\sim$2$\arcmin$ on the sky.
Quoted errors are at the 1-$\sigma$ level, unless otherwise stated.

\section[]{Infrared observations and data reduction}

\subsection{MIPS data and source catalogue}

CLG0218 was observed with \textit{Spitzer} IRAC and MIPS 24$\mu m$ as part of SpUDS\footnote{http://ssc.spitzer.caltech.edu/spitzermission/observingprograms/ legacy/spuds/}. 
The MIPS source catalogue was obtained with the PSF-fitting code StarFinder \citep{Diolaiti} on the publicly available SpUDS images.
Details on the MIPS source extraction and catalogue production can be found in \cite{Tran10}, that presented a subset limited to an area of 
1 Mpc centered on the cluster.
For the present work we select all sources  detected with S/N $>$ 5 (which corresponds to a flux of 40$\mu$Jy) centered on the galaxy ID 16 (see \S 6) that is associated 
with an X-ray point source, out to a projected radius $r$=20$\arcmin$, that corresponds to 10 Mpc at the cluster redshift.

\begin{table}
\caption{Summary of the datasets used in our analysis. The "Type" column refers to spectroscopy (S) or imaging (I).}  
\label{table:3}     
\small
\centering           
\begin{tabular}{llll} 
\hline\hline   
Instrument          & Type    &        Observed Band        &        Field-of-view                \\
\hline                         
IMACS               &  S     &  MOS                                &    15.46$\arcmin \times$15.46$\arcmin$  \\
FMOS                &  S     &  MOS, $1.6-1.8\mu m$       &     30$\arcmin \phi$                \\
MOIRCS            &  S     &  MOS, $0.9-1.7\mu m$       &      7$\arcmin \times$4$\arcmin$    \\
MegaCAM         &  I     &  u                                       &     1 deg$\times$1 deg   \\
SuprimeCAM      &  I     &  $BVRiz$                           &     52$\arcmin \times$52$\arcmin$   \\
WFCAM             &  I     &  J, H, K                              &     41$\arcmin \times$41$\arcmin$   \\
IRAC                &  I     &  3.6/4.5/5.8/8.0  $\mu m$      &     41$\arcmin \times$41$\arcmin$   \\
MIPS                &  I     &  24  $\mu m$                       &     41$\arcmin \times$41$\arcmin$   \\
PACS                &  I     &  100/160 $\mu m$               &     34$\arcmin \times$34$\arcmin$   \\
SPIRE               &  I     &  250/350/500 $\mu m$         &     34$\arcmin \times$34$\arcmin$   \\          
ACIS-S              &  I     &  0.5-10 keV                        &    8.5$\arcmin \times$8.5$\arcmin$  \\
\hline
\end{tabular}
\end{table}

\subsection{Herschel data reduction and photometry}

The \textit{Herschel} 100/160/250/350/500$\mu m$ observations of CLG0218 were carried out as part of several programmes: 
(i) a 50 h guaranteed time programme (GT1, PI Altieri) aimed at studying the star formation history in high redshift (0.8$ < z <$ 2.2) 
galaxy clusters, (ii) 80 h observations of the CANDELS fields (PI M. Dickinson) and (iii) 25 h observations of the HerMES Key Program 
(PI S. Oliver). We note that the PACS observations performed by these different 
programmes cover different areas at different depths.
While the GT1 is centered on a 10$\arcmin \times$10$\arcmin$ region around the cluster centre, 
the deeper CANDELS dataset imaged a larger rectangular area (9$\arcmin \times$20$\arcmin$),
and HerMES imaged a much larger area (30$\arcmin \times$30$\arcmin$) but with shallow exposures.
See in Figure \ref{herschelimage} the co-added image of the 3 programmes.
As a consequence, the sensitivity of the PACS images varies significantly across the FOV, however it is optimal in the cluster region
 (magenta square in Fig. \ref{herschelimage} with 10$\arcmin \times$10$\arcmin$). The 3-$\sigma$ sensitivity ranges from 1.3 (3.9) mJy 
 (GT1+CANDELS+HerMES area) to 4.0 (9.6) mJy (HerMES only) in the 100 (160)$\mu m$ image. 

The PACS \citep{Poglitsch} GT1 observations at 100/160$\mu m$ were acquired on
19 January 2011 (obsids = 1342213032-5) with 4 crossed scan maps
of 2.4h each covering a field of 10$\arcmin \times$10$\arcmin$ and the CANDELS observations
on the Ultra Deep Survey (UDS) field were spread throughout July 2012 for a total of 80 h.
The data were reduced using Hipe 9 \citep{Ott}. 
The data cubes were processed with a standard pipeline where detector
timelines are highpass filtered with a sliding median over 21 readouts
at 100$\mu m$ and 41 readouts at 160$\mu m$ to remove detector drifts and 1/f noise,
with an iterative masking of the sources.

Given the large PSF / beam sizes of PACS and SPIRE, we opt to extract sources using priors, instead of matching blind catalogues 
using nearest neighbour approaches. Using our MIPS catalogue as prior we run HIPE/DAOPHOT to extract the 
sources and perform aperture photometry in the PACS bands.  The aperture photometry is corrected with the encircled energy factors 
given by \cite{Balog} in radii of 6$\arcsec$ at 100$\mu m$ and 9$\arcsec$ at 160$\mu m$. Given the difficulty to obtain reliable errors 
with standard source detection algorithms because of the correlated noise present in PACS data, we compute the photometric errors 
as the 1-$\sigma$ detection limits in each band, in addition to 7\% of the source's flux. 
The astrometry relative to MIPS 24$\mu m$ is excellent and no shift was required to be applied to the PACS maps.

For our analysis we used the SPIRE \citep{Griffin} observations at 250, 350 and 500$\mu m$ from \textit{Herschel}-CANDELS UDS. The \textit{Herschel}-CANDELS UDS observations are one of the deepest SPIRE observations, at a similar depth to GOODS-N from \textit{Herschel}-GOODS \citep{Elbaz11}, achieving instrumental noise much deeper than the nominal SPIRE extragalactic confusion noise \citep{Nguyen}. The \textit{Herschel}-CANDELS UDS observations were performed following an 8-point dithering pattern in order to achieve more homogeneous coverage within the CANDELS UDS area. Adding the shallower smaller area GT1 observations or the HerMES \citep{Oliver} observations in this region does not improve the depth as we are limited by the confusion -- it only creates regions of inhomogeneous coverage within the field. The dithering also allows using a smaller than nominal pixel scale. Although this does not improve the resolution (below the Nyquist sampling), it improves the definition of the sources: the maps were made with (3.6,4.8,7.2)$\arcsec$/pixel for (250,350,500)$\mu m$. The \textit{rms} noise in the three SPIRE bands\footnote{We use the median absolute deviation (MAD) $\times$1.48 as a robust measure of the \textit{rms}; MAD is not affected by the presence of sources in the region.} within the good coverage is (3.3,4.0,5.8)$\times \sim$1.48 mJy at (250,350,500)$\mu m$ and indeed this is similar to the nominal extragalactic confusion noise of (5.8,6.3,6.8) mJy but still within the field-to-field or flux calibration uncertainties.

The SPIRE source detection was performed using Sussextractor \citep[SXT,][] {Smith} with a prior catalogue based on the MIPS+PACS catalogue. SXT finds a maximum likelihood fit of the SPIRE beam at the position of the input catalogue sources in the three bands independently. In each band the quality of the fit (e.g. the signal-to-noise) is a measure of the source 
detection that we also confirmed by visual inspection. When there was no plausible source at the position of the input catalogue we used the $3\sigma$ upper limit in each band, i.e. three times the confusion noise or (17.4,18.9,20.4) mJy at (250,350,500)$\mu m$. 

\section{Ancillary data}

In addition to the MIPS and \textit{Herschel} data we use the optical/near-IR imaging and spectroscopy summarized below 
(see Table 1). Details on the observations and reduction procedures can be found in the appropriate references. 
When referring to the spectroscopic data we intend the reduced source catalogues obtained in the literature or private communications.
We use the optical spectroscopy catalogue of \cite{Papovich} obtained with the IMACS instrument at the Magellan telescope, 
as well as the near-infrared spectroscopy obtained with MOIRCS installed in the Subaru telescope and reported in \cite{Tanaka10}, 
and with FMOS also on Subaru \citep{Tadaki}.

Our optical/near-IR spectral energy distribution (SED) fitting presented below is based on multi-wavelength data collected
from a number of surveys.  Deep $BVRiz$ imaging has been performed with the Suprime-CAM at
Subaru \citep{Furusawa} and we use the publicly available z-band selected catalogue (z$_{AB}$=26.5 mag), which
forms the basis of our catalogue.  We combine this catalogue with the $u$-band photometry taken
with CFHT MegaCAM ($u^*_{AB}(5\sigma, 2"{\rm ap.})=27.0$\footnote{Based on publicly available archival CFHT data.}) by cross-matching objects in the two catalogues
within 1 arcsecond.  We further include the publicly available $JHK$ photometry from DR8plus
taken as part of the UKIRT Infrared Deep Sky Survey (UKIDSS\footnote{http://www.ukidss.org/}, Lawrence et al. 2007).
We use 2$\arcsec$ apertures and apply aperture corrections (which are estimated for each data set)
to measure total magnitudes.  
We note that although the $K$-band would be more appropriate than the z-band to select red (dusty) galaxies at $z$=1.62, the $K$-band
data is 1.5 magnitudes shallower than the z-band ($K_{AB}$=25 mag, 3$\sigma$). 
We performed several checks to assess the z-band versus $K$-band selection using our catalogues and we made SED shape predictions based on the 
limiting magnitudes of the z- and $K$-bands, using the templates of \cite{Berta} shifted to $z$=1.6. We conclude that with the z-band selection we
do not significantly miss MIPS sources (7-8\% of sources at all redshifts) that would otherwise be detected in the $K$-band, therefore we find that the deep z-band 
catalogue is suitable for the galaxy selection, in comparison with the shallower $K$-band catalogue.
The \textit{Spitzer} IRAC observations of CLG0218 obtained by the SpUDS are also included in the optical-NIR SED fitting.

The 85 ksec \textit{Chandra} observation published in \cite{Pierre} is also used here. While no diffuse emission was detected in this 
data, the sub-arcsecond resolution of \textit{Chandra} allowed for the detection of three point-sources associated with confirmed 
cluster members which will be important to evaluate the impact of AGN in the IR SEDs.

\section{Galaxy cluster sample}

As detailed below, our galaxy cluster sample is divided in a more robust, spectroscopically confirmed sample, 
and a photometric redshift sample of galaxies with 24$\mu m$ emission.

\subsection{Spectroscopic galaxies }

Our spectroscopic sample of 46 cluster members originates from 3 observation campaigns: (i) the 
IMACS \citep{Papovich} and (ii) MOIRCS observations \citep{Tanaka10} 
secured 16 cluster members selected by colour and photometric redshifts, and (iii) the more recent large area follow-up with FMOS of [OII] 
narrow-band emitters \citep{Tadaki} independently confirmed 40 cluster members. Of these, ten galaxies overlap 
with the initial IMACS + MOIRCS catalogue. 
A total of ten spectroscopic galaxies extending to a radius of 8.5 Mpc at the cluster redshift have 24$\mu m$ emission and will be used
 in our investigation of the star formation of the cluster and its associated large scale environment.

\begin{table*}
\caption{List of the spectroscopic galaxies at $z$=1.62 (first ten rows) and photo-$z$ candidates with 
[OII] emission (33 rows after break) from Tadaki et al. (2012) detected by MIPS. The infrared properties are based on the 24$\mu m$ 
fluxes and corrected for the mid-IR excess problem following the approach of Rujopakarn et al. (2013). }  
\label{table:2}      
\centering           
\begin{tabular}{ccccccccc} 
\hline\hline           
 ID     & RA         &  DEC        &     z      &   Dist      &    $M_{*}$                                                     &     F24         &   $L_{IR}$                         &   SFR        \\
         &                 &                   &            &   (kpc)   &   ($\times$10$^{10}$ M$_\odot$)   & mJy             &   ($\times$10$^{11}$$L_\odot$)  &  (M$_\odot$/yr)    \\    
\hline              
    6*    &  34.57167  &  -5.18492      &  1.6487       & 915        &  6.35     &  0.0793$\pm$0.00269    &	    5.21$\pm$0.34     &    42.1$\pm$2.9                \\
  7        &  34.59967  &  -5.15572     &  1.6222        &  553       &   2.20    &  0.08455$\pm$0.00340 &	    5.62$\pm$0.37     &    46.9$\pm$3.2                \\
  10     &  34.56325   &  -5.13664     &  1.6224       & 1585      &  4.69     &  0.12074$\pm$0.00379   &    8.64$\pm$0.61     &    72.6$\pm$ 5.2                \\
  16*   &  34.59955   &  -5.17385      &  1.6238       &  0            &   6.72    &  0.28767$\pm$0.00328  &   24.57$\pm$1.96     &   232.2$\pm$19.8               \\
  17   &  34.59786   &  -5.15957        &  1.6248       &   438      &   2.63    &  0.11289$\pm$0.00339   &    7.97$\pm$0.56     &    66.7$\pm$4.7               \\
  22   &  34.62734   &  -5.19286         &  1.6269      &  1027     &   2.99    & 0.07136$\pm$0.00340   &     4.59$\pm$0.30     &    37.9$\pm$2.5                \\
  25   &  34.65436   &  -5.26554         &  1.603         &   3258    &   2.80    &  0.08689$\pm$0.00270  &    5.81$\pm$0.39     &    49.5$\pm$3.3                \\ 
  27   &  34.67099   &  -5.21945         &  1.615         &  2585     &   2.00    &  0.08264$\pm$0.00338  &    5.47$\pm$0.36     &    45.9$\pm$3.1                \\
  60   &  34.32777   &  -5.10499         &   1.599        &  8551     &   1.35    &  0.05534$\pm$0.00300  &    3.38$\pm$0.21     &    28.5$\pm$1.8                \\ 
 137  &  34.69108   &  -5.34808         &   1.599         &   6002    &   1.61   &  0.06622$\pm$0.00376  &    4.19$\pm$0.27     &    34.2$\pm$2.3               \\ 
 \hline  
   1003  &    34.59531  &   -5.08806   &   1.619$\pm$0.098  &  2616 &     28.6     &   0.06221$\pm$0.00336  &      3.89$\pm$0.25     &    32.2$\pm$2.1        \\        
  1005  &    34.70046  &   -5.1923     &   1.432$\pm$0.086  &  3132 &     2.27  	 &  0.09774$\pm$0.00288  &    6.70$\pm$0.46     &    56.1$\pm$3.9        \\
  1008  &    34.49514  &   -5.33445   &   1.747$\pm$0.089  &  5844 &     5.16  	 &   0.27709$\pm$0.00343  &   23.48$\pm$1.86   &  222.3$\pm$18.8        \\
  1014  &    34.70506  &   -5.02685   &   1.618$\pm$0.096   &  5511 &     1.33     &  0.04944$\pm$0.00323  &       2.94$\pm$0.18   &    24.3 $\pm$1.5        \\
  1015  &    34.65158  &   -5.26029   &   1.703$\pm$0.048  &  3084 &      3.08  	 &  0.08388$\pm$0.00282  &    5.57$\pm$0.37    &    46.5$\pm$3.1         \\
  1016  &    34.6953    &   -5.1313     &   1.529$\pm$0.053  &  3200 &     2.65       &   0.11576$\pm$0.00322 &     8.21$\pm$0.57    &    69.1$\pm$4.9         \\
  1046  &    34.5648    &   -5.11625   &   1.539$\pm$0.044  &  2052 &     2.42   	 &   0.11944$\pm$0.00339  &   8.53$\pm$0.60    &    71.8$\pm$5.1         \\
  1053  &    34.443      &   -5.38437   &  1.584$\pm$0.105   &  8003 &     1.99   	 &  0.04626$\pm$0.00291  &    2.72$\pm$0.16    &    22.4$\pm$1.4         \\
  1059  &    34.77112   &  -5.1696     &   1.651$\pm$0.05  &   5238 &    7.34         &   0.62738$\pm$0.00435 &    62.81$\pm$5.55   &   636.5$\pm$60.0        \\
  1078  &    34.48349   &  -5.07822   &   1.609$\pm$0.111   &  4584 &    2.37      &   0.06652$\pm$0.00347  &      4.21$\pm$0.27   &    35.0$\pm$2.3         \\
  1079  &    34.7666    &   -5.09876   &   1.585$\pm$0.064   &  5588 &    3.11   	&   0.11590$\pm$0.00330  &    8.22$\pm$0.57   &    69.2$\pm$4.9         \\
  1080  &    34.67669   &  -5.39654    &   1.759$\pm$0.097  &  7194 &    3.69   	&   0.07001$\pm$0.00332 &     4.48$\pm$0.29   &    37.3$\pm$2.4         \\
  1088  &    34.63499   &  -5.14839    &  1.557$\pm$0.054   &  1334 &    1.11     &    0.07281$\pm$0.00330 &       4.70$\pm$0.30   &    39.1$\pm$2.6         \\
  1109  &    34.41346   &  -5.23934    &  1.420$\pm$0.074    &  6018 &   4.19     &   0.12107$\pm$0.00343 &        8.68$\pm$0.61   &   73.0$\pm$5.2         \\
  1111  &    34.47474   &  -5.18248    &  1.563$\pm$0.106   &  3816 &    2.78     &   0.09399$\pm$0.00400 &        6.39$\pm$0.43   &   53.5$\pm$3.7          \\
  1122  &    34.59218   &  -5.06676   &   1.569$\pm$0.041    &  3265 &  1.65      &   0.06509$\pm$0.00342  &       4.11$\pm$0.26   &   34.1$\pm$2.2          \\
  1123  &    34.54131   &  -5.10883   &   1.447$\pm$0.071   &   2662 &    1.84    &  0.07649$\pm$0.00282  &        4.99$\pm$0.33   &   41.5$\pm$2.8          \\
  1125  &    34.62784   &  -5.4367     &   1.555$\pm$0.040   &   8066 &   1.66     &   0.06283$\pm$0.00344 &        3.94$\pm$0.25   &   32.6$\pm$2.1          \\
  1155  &    34.54086   &  -5.12823   &   1.607$\pm$0.200   &   2259 &    1.77    &  0.08780$\pm$0.00391   &       5.89$\pm$0.39   &   49.2$\pm$3.4          \\
  1164  &    34.55123   &  -5.1129     &   1.525$\pm$0.087     &   2367 &     2.71 &   0.11608$\pm$0.00327   &      8.24$\pm$0.58   &   69.3$\pm$4.9          \\
   1169  &    34.72103   &  -5.17092   &   1.494$\pm$0.063   &   3708 &    11.5   &   0.28355$\pm$0.00334   &     24.14$\pm$1.92   &  229.0$\pm$19.5         \\
  1176  &    34.49677   &  -5.18354   &   1.629$\pm$0.121   &   3150  &    2.32    &  0.08359$\pm$0.00338   &      5.55$\pm$0.37   &   46.3$\pm$3.1           \\
  1202  &    34.45989   &  -5.14504   &   1.533$\pm$0.072   &   4348  &    2.83    &  0.08113$\pm$0.00352   &      5.35$\pm$0.35   &   44.7$\pm$3.0           \\
  1213  &    34.53631   &  -5.39567   &   1.51$\pm$0.059     &   7046  &    1.57    &  0.06741$\pm$0.00277   &     4.28$\pm$0.27   &   35.6$\pm$2.3            \\
  1217  &    34.73699   &  -5.23179   &    1.61$\pm$0.093   &   4554  &    1.36  	&    0.05656$\pm$0.00344  &   3.47$\pm$0.21   &   28.7$\pm$1.8            \\
  1225  &    34.70364   &  -5.28452   &   1.496$\pm$0.095    &   4636  &    1.87  &   0.06872$\pm$0.00338   &     4.38$\pm$0.28   &   36.4$\pm$2.4            \\
  1226  &    34.70381   &  -5.2425     &   1.617$\pm$0.179   &    3809 &    3.21   &   0.10240$\pm$0.00400   &     7.08$\pm$0.49   &    59.4$\pm$4.1            \\
  1228  &    34.37853   &  -5.41709   &   1.495$\pm$0.074    &  10019 &    2.0    &  0.07896$\pm$0.00314   &      5.18$\pm$0.34   &    43.2$\pm$2.9            \\
  1240  &    34.77837   &  -5.10404   &   1.57$\pm$0.072     &   5856 &    1.86  	&  0.05682$\pm$0.00314  &    3.49$\pm$0.22   &    28.8$\pm$1.8            \\
  1260  &    34.60447   &  -5.26243   &   1.548$\pm$0.024    &   2708 &   1.16    &  0.08142$\pm$0.00276  &       5.38$\pm$0.36   &    44.9$\pm$3.0            \\
  1272  &    34.75204   &  -5.21685   &   1.572$\pm$0.065    &   4829 &     0.90   &  0.05142$\pm$0.00309   &     3.09$\pm$0.19   &    25.5$\pm$1.6            \\
  1312  &    34.61321   &  -5.04895   &   1.526$\pm$0.090    &   3823 &     3.73 	 &  0.15646$\pm$0.00453  &  11.80$\pm$0.86   &  100.1$\pm$10.9            \\
  1313  &    34.60363   &  -5.25612   &   1.662$\pm$0.221    &   2521 &     5.37    & 0.06744$\pm$0.00276  &      4.28$\pm$0.27   &   35.6$\pm$2.3             \\
   \hline  
\end{tabular}
   \flushleft  \hspace{0.3cm}   * X-ray emission
\end{table*}

\subsection{Photo-$z$ members} 

We compute photometric redshifts ($z_{phot}$ hereafter) using the $uBVRIZJHK(3.6,4.5)\mu m$ catalogue
constructed above.
We use the SED fitting code described in \cite{Tanaka13}.  In short, we construct
model templates using an updated version of the \cite{Bruzual} models, which includes
an improved treatment of the thermally pulsating AGB stars.  We assume exponentially decaying
star formation histories with the decay time scale allowed to vary.  Dust attenuation is included
using the attenuation curve by \cite{Calzetti2000}.  Emission lines are added to the spectra
using the emission line intensity ratios given in \cite{Inoue} assuming the \cite{Calzetti} attenuation to the emission lines.  
We apply the same template error function as in \cite{Tanaka13}
in order to reduce systematic differences between observations and model templates.
We infer the galaxies physical parameters such as stellar mass and SFR from the fitting of the stellar population synthesis models. 
Each model parameter is marginalized over all the other parameters and
we take the median of the probability distribution as a point estimate and quote the 68\% interval as an uncertainty.

\begin{figure}
\includegraphics[width=7.5cm,angle=0]{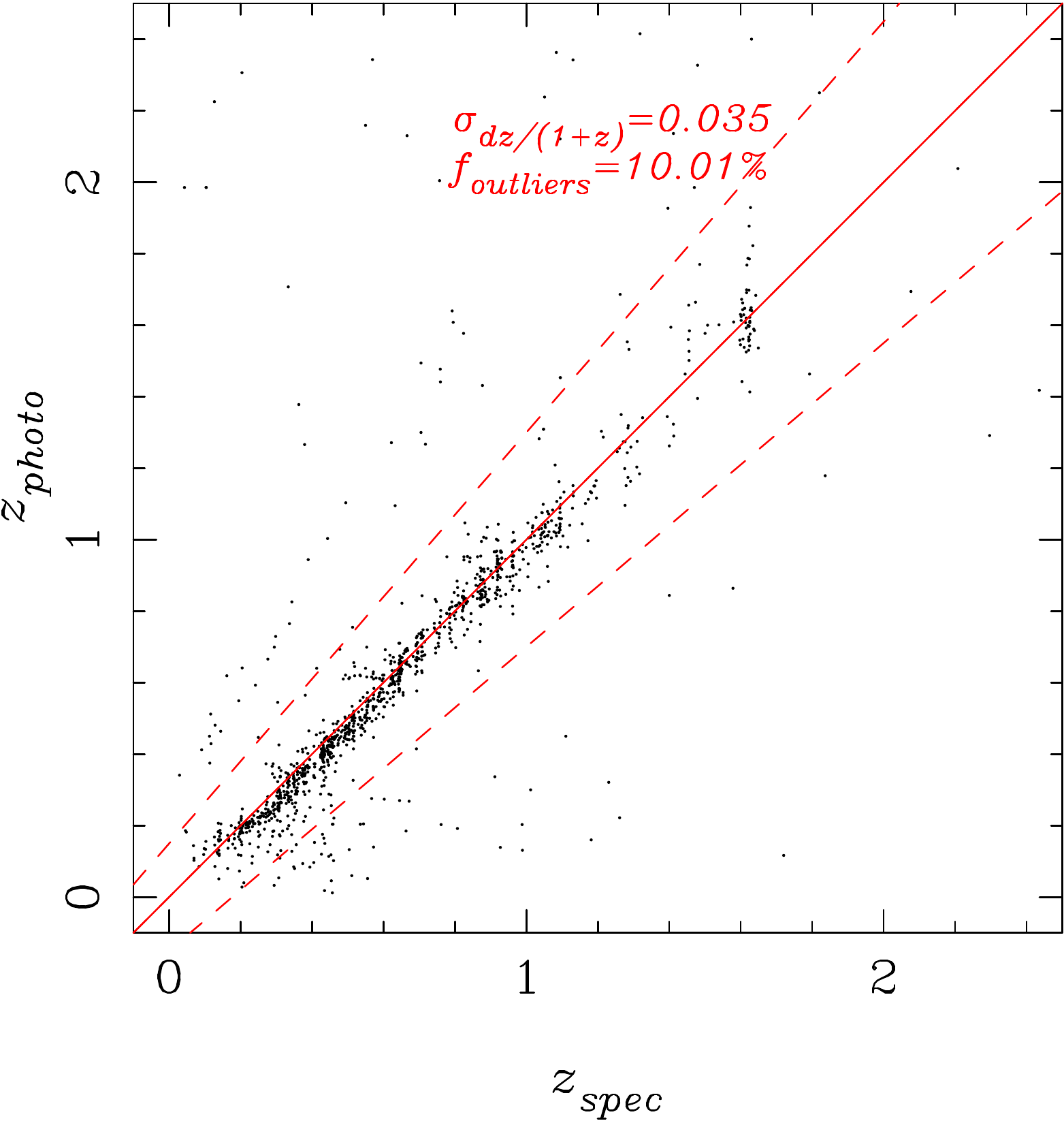} 
 \caption{ Comparison between the spectroscopic and photometric redshifts for objects in the area of our analysis. The solid line corresponds to 
 $z_{phot} = z_{spec}$, the dashed lines are $|z_{phot} - z_{spec}|$/(1+$z_{spec}$) = $\pm$0.15.  Objects outside of these dashed lines are regarded as outliers.  
 The scatter is also estimated with a standard estimator of $\sigma$ = 1.48 MAD (($z_{phot} - z_{spec}$)/(1+$z_{spec}$)) where MAD 
 is median absolute deviation and the factor 1.48 is a conversion factor from MAD to sigma for the normal distribution. }
 \label{photoz}
\end{figure}

We compare the resultant $z_{phot}$ with spectroscopic redshifts from the literature
\citep{Smail, Papovich,Tanaka10, Simpson12, Tadaki}; Akiyama et al. in prep.).  The redshift catalogue is a collection of follow-up campaigns targeting
different types of objects and is quite heterogeneous therefore we have to exercise some caution when interpreting the numbers. 
We find $\sigma(\Delta z/(1+z))\sim0.04$ with an outlier rate of $\sim10\%$ (see Fig. \ref{photoz}).
The 1-$\sigma$ range of the photo-z uncertainty at $z=1.62$ is thus $\sigma (z_{phot}) = 0.04 \times (1+z) = 0.10$.
The $\pm 2 \sigma$ range in photometric redshifts to select candidate galaxy cluster members is then 1.42$< z_{phot} <$1.82.

Using a nearest neighbour algorithm with a distance of 1.5$\arcsec$ we match the MIPS sources with our photo-$z$ catalogue and we
find 981 24$\mu m$ sources within an radius of 10 Mpc projected distance from the cluster center with a photo-$z$ 1.42--1.82.
However, given our large number of sources we can afford to apply a stricter $z_{phot}$ cut allowing for a 1 $\sigma$ range, ie., 1.52--1.72. 
With this cut we have 679 MIPS sources in that same area.
Constraining our analysis to the inner 1 Mpc projected radius we have 14 MIPS photo-$z$ members, of which 4 
are spectroscopic members.

In addition, we use in our study the full catalogue with 352 narrow-band emitters of which 40 were 
spectroscopically confirmed using Subaru/FMOS (see above). For the remaining sources we cross-match the photo-$z$ catalogue, 
and we obtain photometric redshifts for 308 emitters, of which 269 OII have 1.42$< z_{phot} <$ 1.82 (we relax the constraints on the 
photometric range because the indication of [OII] emission strengthens the hypothesis of these galaxies having a redshift consistent with 
the cluster). Thirty-three [OII] members are associated with MIPS detections, distributed in a radial range 0--8 Mpc.

In summary, in the following sections we will investigate the mid-to-far infrared properties of 
(i) 10 spectroscopic galaxies, (ii) 33 OII emitters with 1.42$<z_{phot}<$1.82 and (iii) 679 photo-$z$ members with a stricter 
redshift range of 1.52--1.72. These 693 galaxies are spread around a radius of 10 Mpc from the cluster center which allows us to study 
the large scale environment around CLG0218. Specifically, there are 14 galaxies at $r<$ 1 Mpc, 60 at 1$<r<$3 Mpc and 619 at 3$<r<$10 Mpc. 

\section{The total infrared luminosity, $L_{IR}$, and the derived SFRs}

\subsection{Results from MIPS only}

The most commonly used infrared SED template libraries to describe the spectra of star-forming galaxies 
are those of \cite{Chary}, \cite{Dale} and \cite{Rieke}. 
\textit{Herschel} observations showed that using these libraries to infer the total infrared luminosity from 
the 24$\mu m$ flux alone significantly overestimates the $L_{IR}$ by a factor 2--7 for galaxies
at $z>$1.5 with $L_{IR}$ $>$10$^{10} L_\odot$ \citep[e.g.,][]{Nordon, Rujopakarn}.

To overcome this so-called mid-IR problem, \cite{Rujopakarn13} recently derived an empirical \textit{stretching factor} for the 
\cite{Rieke} SED templates, 
based on the evidence that IR luminous star-forming galaxies at 1 $< z <$ 3 have extended star-forming regions, 
as opposed to the strongly nuclear concentrated starbursts in local LIRGs and ULIRGs (note that Rieke et al. 2009 consider $L_{IR}$ from 5--1000$\mu m$). 
We use this approach here to obtain the total infrared luminosities. In the next section 
we will also check this method by comparing these values with the ones computed with both MIPS (corrected) and \textit{Herschel}.  

To convert the luminosity to a SFR we apply the Kennicutt relation \citep{Kennicutt} modified for the Chabrier IMF 
\citep{Chabrier, Erb}:
\begin{equation}
  SFR\,(M_\odot yr^{-1})= 10^{-10} L_{IR}/L_\odot
\end{equation}

\noindent For simplicity we assume a redshift of 1.62 for all galaxies in these computations and we estimate the errors on $L_{IR}$ and SFR by considering 
photometric redshift errors of 0.1. The star formation rates range from 18 to 2500 M$_\odot$/yr, with a median of 55 M$_\odot$/yr.
 
\begin{table*}
\caption{Infrared fluxes (24--500 $\mu m$) of the spectroscopic and photo-$z$ galaxies at $z \sim$1.62 with \textit{Herschel} detection. 
All fluxes are in mJy and the sources are ordered in projected cluster centric distance. The top, middle and bottom sections refer to the 
galaxies at $r<$ 1 Mpc, at 1$< r <$3 Mpc and at $r>$3 Mpc, respectively. } 
\label{table:2}      
\centering           
\begin{tabular}{lllllll} 
\hline\hline          
ID     &   F$_{24}$  &   F$_{100}$    & F$_{160}$ & F$_{250}$   &    F$_{350}$   &  F$_{500}$   \\     
\hline              
   16      &   0.288$\pm$0.003  &   8.6$\pm$1.0  &   17.8$\pm$2.5   &   22.8$\pm$6.0   &   19.3$\pm$6.5   &   14.9$\pm$7.0      \\        
 17742  &   0.083$\pm$0.004   &  --       &  10.9$\pm$2.1   &  22.6$\pm$6.0   &  27.6$\pm$6.6   &   26.4$\pm$7.1    \\
  4862  &  0.214$\pm$0.003      &   7.6$\pm$1.0 &  22.1$\pm$2.8     &   34.4$\pm$6.2   &  40.4$\pm$6.8   &    27.2$\pm$7.0      \\
\hline
 3509  &  0.259$\pm$0.003    &  12.1$\pm$1.3  &   16.5$\pm$1.5    &   20.3$\pm$6.0   &   13.7$\pm$6.4   &     --                     \\
  3622  & 0.255$\pm$0.004     &  11.6$\pm$1.2  &   26.9$\pm$3.2     &   43.0$\pm$6.4   &  40.8$\pm$6.8  &     23.7$\pm$7.0    \\
  9896  &  0.136$\pm$0.004      &    8.6$\pm$1.0  &  15.2$\pm$2.4   &  22.7$\pm$6.0   &  21.6$\pm$6.5   &   15.6$\pm$7.0     \\
 5238  &  0.205$\pm$0.004      &   2.5$\pm$0.6  &   8.3$\pm$1.9     &  21.0$\pm$6.0   &  25.5$\pm$6.5   &     21.0$\pm$7.0     \\
  2154  &   0.339$\pm$0.004   &  19.8$\pm$1.8  &  34.8$\pm$3.7   &    49.1$\pm$6.5   &   33.0$\pm$6.6    &    --                      \\
  2480  &   0.312$\pm$0.002   &  --                      &   8.6$\pm$1.9     &    24.6$\pm$6.0   &   32.4$\pm$6.6  &    24.1$\pm$7.2    \\
\hline
  2787  &  0.292$\pm$0.003    &  --                       &    7.7$\pm$1.8   &  18.4$\pm$6.0    &   21.2$\pm$6.5  &     15.1$\pm$7.0     \\
 12996  &   0.110$\pm$0.004   &    3.7$\pm$0.7   &   4.7$\pm$1.6   & 17.7$\pm$6.0    &   24.2$\pm$6.5  &  24.3$\pm$7.0 \\
  3270  &  0.268$\pm$0.004    &  5.3$\pm$0.8     &    18.4$\pm$2.6  &   44.9$\pm$6.4   &   42.7$\pm$6.8   &     35.4$\pm$7.2    \\
   5596  &   0.197$\pm$0.004     &   3.7$\pm$0.7  &   14.1$\pm$2.3   &  20.8$\pm$6.0     &  16.9$\pm$6.5   &   --                   \\
  1614  &   0.402$\pm$0.003   &  13.3$\pm$1.4  &  31.9$\pm$3.5  &   58.4$\pm$6.7   &  50.7$\pm$7.0      &    30.5$\pm$7.1     \\
  2664  &   0.299$\pm$0.003   &  16.1$\pm$1.5   &   39.7$\pm$4.1  &    54.9$\pm$6.6   &   35.8$\pm$6.7    &    16.2$\pm$7.0     \\
  1608  &   0.403$\pm$0.004   &  10.4$\pm$1.1  & 18.1$\pm$2.6   &  26.7$\pm$6.1   &   24.3$\pm$6.5      &   --      \\  
  4234  & 0.233$\pm$0.004      &  4.8$\pm$0.8   &   10.3$\pm$2.0     &  19.1$\pm$6.0    & 20.8$\pm$6.5   &    16.7$\pm$7.0   \\
  9772  &  0.137$\pm$0.004      &   1.6$\pm$0.5   &  12.4$\pm$2.2  & 30.5$\pm$6.2   &  33.7$\pm$6.7     &   35.4$\pm$7.2     \\  
  9079  &  0.144$\pm$0.004      &   1.2$\pm$0.5   &  12.3$\pm$2.1  & 25.8$\pm$6.1   &  28.6$\pm$6.6     &    40.3$\pm$7.3     \\  
  1393  &   0.435$\pm$0.004   &   8.3$\pm$1.0  &  20.2$\pm$2.7   &   35.7$\pm$6.2   &   33.6$\pm$6.7     &   23.3$\pm$7.0     \\
  1829  &   0.372$\pm$0.003   &  8.5$\pm$1.0   &   20.5$\pm$2.7   &   38.6$\pm$6.3   &  31.0$\pm$6.6     &    17.8$\pm$7.0     \\
  5890  &   0.191$\pm$0.003     &   1.5$\pm$0.5  &   6.0$\pm$1.7    &   17.4$\pm$6.0    & 23.9$\pm$6.6    &    19.1$\pm$7.0       \\ 
  4877  &  0.214$\pm$0.003      &   8.0$\pm$1.0  &  18.4$\pm$2.6    &  32.6$\pm$6.2    &  27.5$\pm$6.6   &    --      \\ 
  1439  &   0.427$\pm$0.004   &   7.7$\pm$1.0  &  26.6$\pm$3.2   &   39.4$\pm$6.2   &   36.5$\pm$6.7     &  22.2$\pm$7.0       \\
 4626  &  0.220$\pm$0.003     &  1.4$\pm$0.5   &   7.5$\pm$1.8       &  21.9$\pm$6.0    & 22.8$\pm$6.5   &   15.7$\pm$7.0      \\
  7053  &   0.171$\pm$0.004     &   8.4$\pm$1.0   &  20.1$\pm$2.7  &   31.8$\pm$6.1    &  25.6$\pm$6.5  &     15.1$\pm$6.9    \\
  14261  &   0.102$\pm$0.003   &    3.8$\pm$0.7  &   6.6$\pm$1.8   &  25.6$\pm$6.1  &  29.1$\pm$6.6  &  28.6$\pm$7.1 \\      
   10896  &   0.127$\pm$0.004   &    10.9$\pm$1.1  &  13.4$\pm$2.2  &  22.1$\pm$6.0   &  15.2$\pm$6.4   &   --      \\
  2641  &   0.301$\pm$0.004   &  1.9$\pm$0.6    &   11.2$\pm$2.1   &    17.5$\pm$6.0   &    16.0$\pm$6.4   &    --                      \\
   \hline  
\end{tabular}
\end{table*}

\subsection{Results from MIPS+Herschel detections}

\begin{figure}
\includegraphics[width=8.5cm,angle=0]{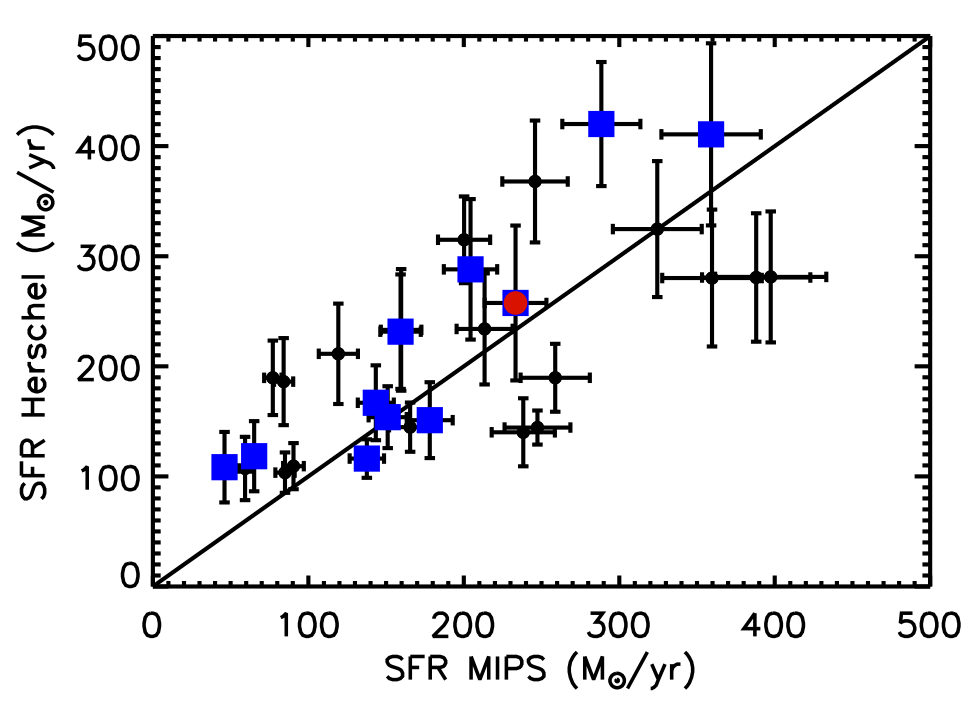} 
 \caption{ SFR(MIPS) vs SFR(\textit{Herschel}). The red dot indicates the brightest infrared cluster galaxy (ID 16). The blue squares indicate possible 
 contamination from neighbouring galaxies that could boost the SFRs obtained with \textit{Herschel}.}
 \label{sfrcomparison}
\end{figure}

Even though our infrared analysis is driven by the MIPS data because the sensitivity and spatial resolution of MIPS enables the detection and characterization of a large galaxy 
sample, we have \textit{Herschel} detections for a smaller yet sizable subset of the MIPS catalogue. This is important because it will allow us to evaluate 
the reliability of the 24$\mu m$-derived SFRs.
Of the 693 galaxies at $z\sim$1.62 individually detected with MIPS, about half are not covered by PACS. Although 97 sources have 
3-$\sigma$ detections in both PACS bands, we find it more reliable for the FIR SED fitting to have also a measurement (not upper limit) of 
the 250$\mu m$ band, given that the peak of the FIR SED is around the 250$\mu m$ at this redshift. 
 A subset of 40 galaxies have 3-$\sigma$ fluxes in 3 \textit{Herschel} bands, either the 100/160/250$\mu m$ bands or the 160/250/350$\mu m$ bands. 
After visual inspection for contamination by close neighbours and reliability of the detections - particularly those which lie in the PACS area with 
poorer sensitivity - there are 29 $z_{phot}$ members with a robust FIR SED fit. 
Using the \textit{Spitzer} and the \textit{Herschel} flux measurements, we fit the galaxy SEDs with  
{\tt LePhare} \citep{Arnouts, Ilbert}, that adjusts SED templates from the chosen \cite{Chary} library
based on a $\chi^{2}$ minimization. In Figure \ref{sfrcomparison} we compare the star formation rates obtained by direct integration of the FIR SED 
anchored on the \textit{Herschel} data with the ones extrapolated from the MIPS 24$\mu m$ fluxes and corrected according to \S 5.1.
 For the overlapping sample of 29 galaxies, the range of SFR(24$\mu m$) is  46 -- 397, with a median of 178 M$_\odot$/yr.  
Even though there is some scatter around the 1--1 relation, we find a strong correlation between these 2 
measures of SFR, quantified by a Spearman rank $\rho$=0.72, with a probability for a null-hypothesis of 9.3$\times$10$^{-05}$.
 
 It is hard if not impossible to investigate whether the scatter is due to measurement errors or uncertainties in the calibration of the SFRs.
 For most galaxies, the \textit{Herschel} fluxes are close to the 3-$\sigma$ detection limit, with SFR errors ranging from 11 to 30\% with a median of 21\%, 
 hence we do not have enough information to identify the origin of the scatter. Nevertheless, visual inspection of the K-band shows that a third of this 
 sample has close neighbours (i.e., within $r$=6$\arcsec$), hence contamination may be an issue causing an artificial boost to the star formation rates derived 
 from \textit{Herschel} fluxes. Indeed, as evidenced in Fig. \ref{sfrcomparison}, this may explain why some sources have SFRs(Herschel) significantly 
 greater than the ones measured from 24$\mu m$.

\subsection{Dust extinction}

The mid-to-far infrared data allow us to obtain a robust estimate of the amount of extinction due to dust by directly 
comparing the star formation-rates obtained from the [OII] luminosity with the ones measured from the 24$\mu m$ fluxes 
(or full FIR SED if we have \textit{Herschel} detections). 

\begin{figure}
\hspace{0.4cm}\includegraphics[width=6.5cm,angle=0]{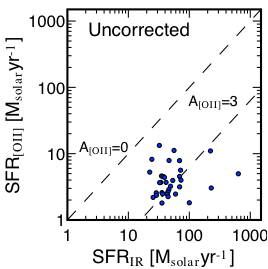} 
 \caption{Comparison of the SFR([OII]) vs SFR(MIPS) for a subset of 33 galaxies detected both in narrow band imaging and MIPS. 
 The diagonal dash lines show the amount of extinction in [OII] necessary to reconcile the two SFR diagnostics. }
 \label{oii}
\end{figure}

Using equation (4) of \cite{Kewley}, we compute dust-uncorrected SFR([OII]) for a sample of 33 cluster galaxies detected 
both in narrow band imaging and MIPS. We find that, in order to reconcile the SFR([OII]) with 
the ones obtained using the 24$\mu m$ observations (corrected for the mid-IR excess problem) we require A([OII])$\sim$3 magnitudes 
(see Fig. \ref{oii}). Similar extinction levels have been reported in the literature for galaxies at $z\sim$1.5 and with similar stellar mass \citep{Koyama11, Kashino}.
This value is a factor 3 higher than the average extinctions we obtain with the optical/NIR SED fitting analysis.
The narrow range of stellar masses (median of 4$\times$10$^{10}$ M$_\odot$) of this sample does not enable the investigation of the role of $M_{*}$ with 
extinction.

\section{The brightest FIR cluster galaxy}

\begin{figure*}
\includegraphics[width=13.5cm,angle=0]{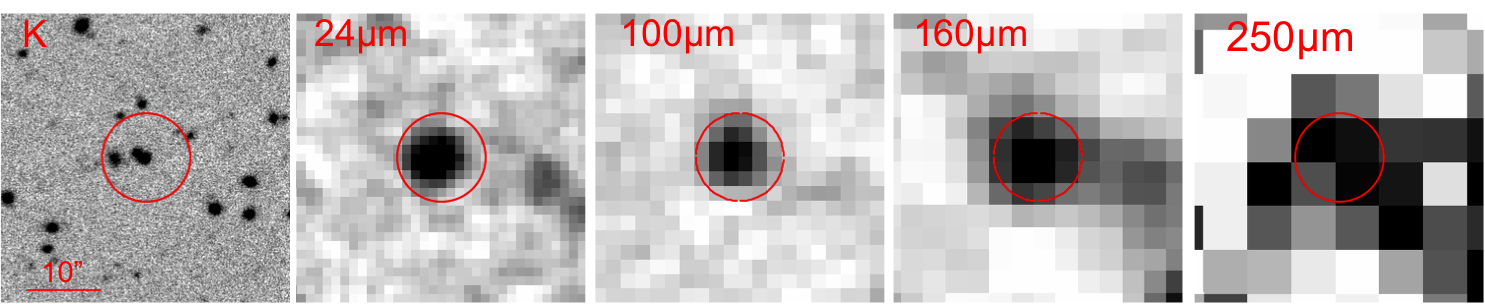} 
 \caption{The brightest infrared cluster galaxy (ID 16). 40$\arcsec \times$40$\arcsec$ cutouts centered on the galaxy, from left-to-right: 
 K-band, 24$\mu m$, 100$\mu m$, 160$\mu m$, 250$\mu m$. }
 \label{ima16}
\end{figure*}

\begin{figure}
\includegraphics[width=8.5cm,angle=0]{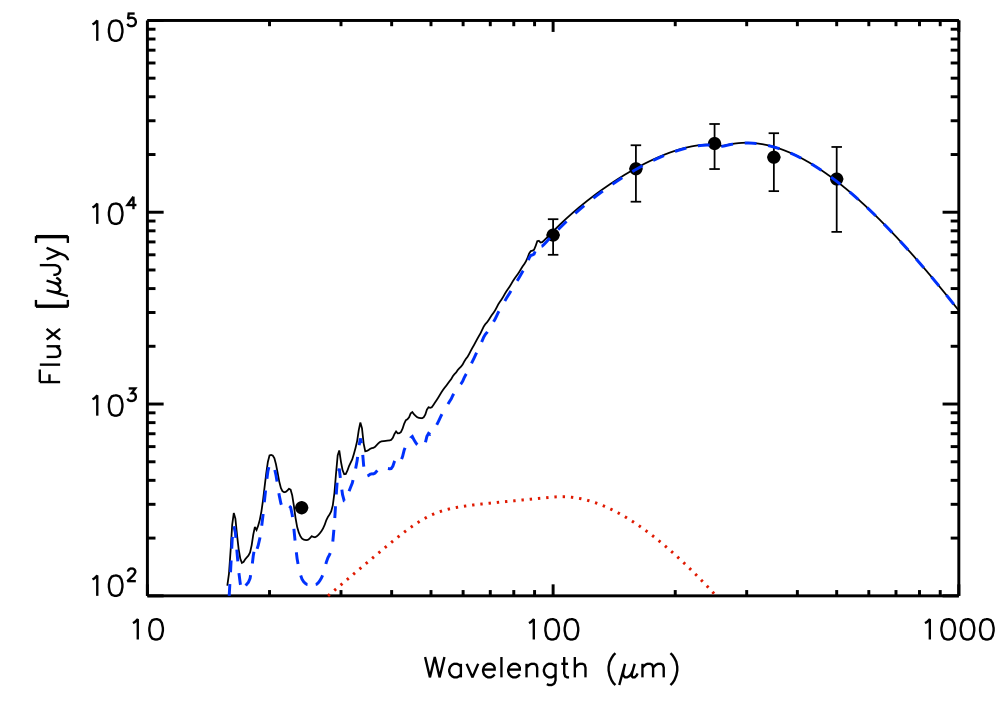} 
 \caption{Spectral energy distribution (black points) of the brightest infrared cluster galaxy (ID 16) and best-fit (blue solid line). The negligible 
 contribution from the AGN component is shown by the red dotted line. }
 \label{sed16}
\end{figure}

CLG0218 hosts a relatively massive (7$\times10^{10}$ M$\odot$) galaxy at the spectroscopic redshift 1.6238 (ID 16 in our catalogue) that 
dominates the cluster (i.e., $r<$ 1 Mpc) infrared emission. In previous work \citep[e.g.,][] {Tran10, Bassett} this galaxy was taken as the cluster centre hence 
we adopt this approach here too. 
Following the procedure described in \S5.2 we obtain a total infrared luminosity $L_{IR, Herschel}$=3.3$\pm$1.0$\times10^{12} L_\odot$, confirming that
ID 16 is a ULIRG, and SFR(Herschel) = 256$\pm$70 M$\odot$/yr. We also estimate $L_{IR}$ using the method previously described based on the 
extrapolation from 24$\mu m$, and correct it for the mid-IR excess problem. We note that \cite{Rujopakarn13} stress that the proposed 
methodology may not apply to galaxies hosting an AGN. 
We also note that for $z>$1 the new 24$\mu m$ indicator underestimates by a factor up to 0.5 dex the total infrared luminosity 
in comparison with \textit{Herschel} derived luminosities \citep[Fig. 3 of][] {Rujopakarn13}.
As expected, for ID 16 we find a lower SFR from the 24$\mu m$ flux,  SFR(24$\mu m$) = 232$\pm$20 M$_\odot$/yr, although this is fully 
consistent with the \textit{Herschel} derived value.
This galaxy is also associated with X-ray emission detected by \textit{Chandra}, therefore
we investigate the contribution of the AGN component to the FIR emission, using the programme {\tt DecompIR} \citep{Mullaney}, 
an SED model fitting software that attempts to separate the AGN from the host star forming (SF) galaxy.
Briefly, the AGN component is an empirical model based on observations of moderate-luminosity local AGNs, whereas the
5 starburst models were developed to represent a typical range of SED types, with an extrapolation beyond 100$\mu m$ 
using a grey body with emissivity $\beta$ fixed to 1.5. For more details see \cite{Mullaney} and \cite{Seymour}.
The best-fit model obtained with {\tt DecompIR}, 'SB5', indicates that the host galaxy dominates the FIR emission, 
with a negligible ($\sim$4\%) contribution from the AGN to the total luminosity as shown in Figure \ref{sed16}.

Since we have a good sampling of the FIR SED for this galaxy we are able to measure the galaxy dust temperature using a 
modified black body model with an emissivity index $\beta$=1.5. The fit relies only on the PACS and SPIRE data that effectively 
straddle the peak of the SED.  We find that the cold dust component 
of the galaxy has a temperature $T_{dust}$=34.5$\pm$4.2 K, a value typical of high-$z$ ULIRGs \citep[e.g.,][] {Hwang}.

We note that, if any, the contamination from the close neighbour located at $\sim$1.5$\arcsec$ from ID 16 must be small since 
it is not centered with the 24$\mu m$ emission (see Fig. \ref{ima16}). This galaxy may belong to the cluster, with $z_{phot}$=2.0.

\begin{figure}
\hspace{0.4cm} \includegraphics[width=7.cm,angle=0]{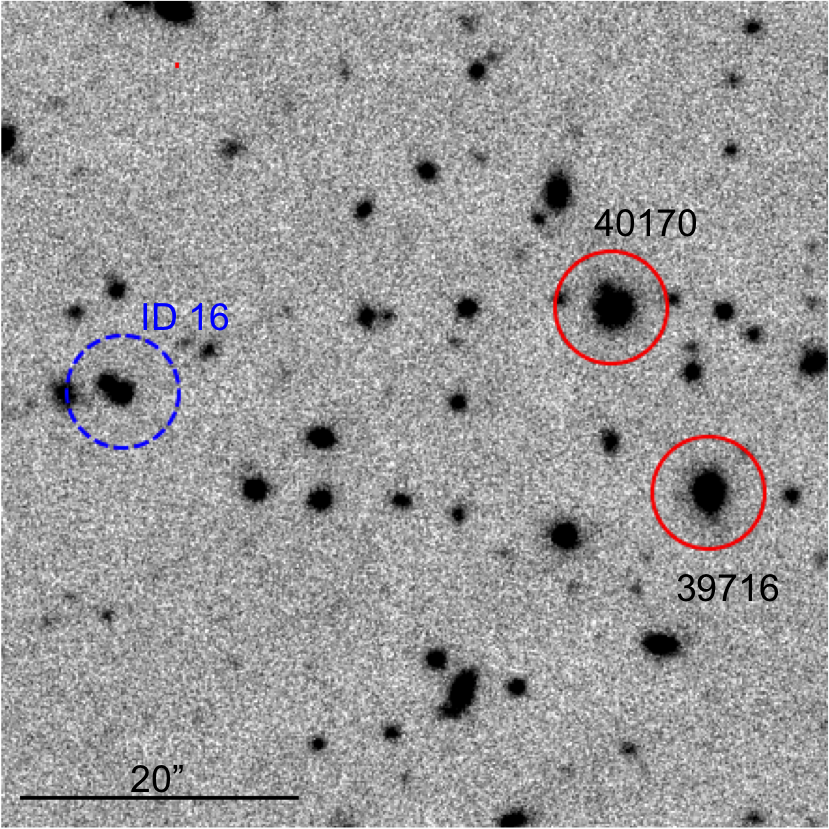} 
 \caption{The cluster CLG0218 seen in the K-band (1$\arcmin \times$1$\arcmin$): ID 16 and the two most massive (quiescent) cluster galaxies. 
 The angular scale of 20$\arcsec$ shown in the figure corresponds to a physical size of 170 kpc at redshift $z$=1.62.  }
 \label{prebcg}
\end{figure}

Since ID 16 clearly dominates the cluster SFR (the second highest star forming galaxy in the cluster has SFR=160 M$_\odot$/yr) 
we can draw a comparison with the Spiderweb galaxy (PKS1138) embedded in a proto-cluster system 
at $z$=2.16 \citep{Miley}, because less than one Gyr separates the 2 galaxies in the adopted cosmology. 
Unlike the Spiderweb, ID16 is not associated with significant radio emission as seen by the deep VLA catalogue of \cite{Simpson}, 
nor does the AGN (X-ray) emission appear to significantly contribute to the FIR SED.
The recent \textit{Herschel} study of PKS1138 by \cite{Seymour} shows that 60\% of the total infrared luminosity is due to the AGN 
component, and the star formation rate from the starburst corresponds to 772$\pm$83 M$_\odot$/yr (we scaled here the published 
value to the Chabrier IMF). Therefore, the core of CLG0218 appears to have a different history in comparison to PKS1138. 
Studies on the formation of galaxy clusters indicate that the development of the intracluster medium in the initial process of  cluster virialization 
takes place around the brightest cluster galaxy \cite[BCG, e.g.,][]{Voit}. Therefore, another interesting aspect to explore is whether galaxy ID 16 could be seen as a 
precursor of the cluster BCG \citep{Lapi}. Again in PKS1138, the Spiderweb galaxy is undoubtedly the dominant galaxy of the system, with a stellar mass 
of the order of 10$^{12}$ M$_\odot$ and will most likely form the brightest cluster galaxy. 
In CLG0218 the scenario is different. In \cite{Papovich11} the formation of the BCG is discussed using a sample of 
$K$-band selected quiescent galaxies. The authors consider that the two most massive ($\sim$2$\times$10$^{11}$M$_\odot$) 
spectroscopically confirmed galaxies (ID 39716 and ID 40170 in their nomenclature), separated by a projected distance of 126 kpc, will eventually merge 
and form the BCG with a stellar mass $>$3$\times$10$^{11}$ M$\odot$. Here we raise the possibility that ID 16, though at a projected distance of 
about 300 kpc from these two massive, quiescent galaxies, could also merge with them, contributing to the BCG as well (see Fig. \ref{prebcg}).
With the current data this is difficult to assess let alone prove, mainly because it's not possible to accurately establish the cluster centre, 
however it would be interesting to follow-up on this point with kinematic studies.

\section{The sSFR--$M_{*}$ relation}

In this section we investigate the relation between stellar mass and star formation rate. 
The stellar masses were computed with the same optical/NIR SED fitting technique as described in section 4.2. 
We estimate a 5$\sigma$ mass completeness limit of 1.5$\times$10$^{10}$ M$_\odot$ based on the z-band limiting magnitude (26.5 mag).

\begin{figure*}
\includegraphics[width=8.5cm,angle=0]{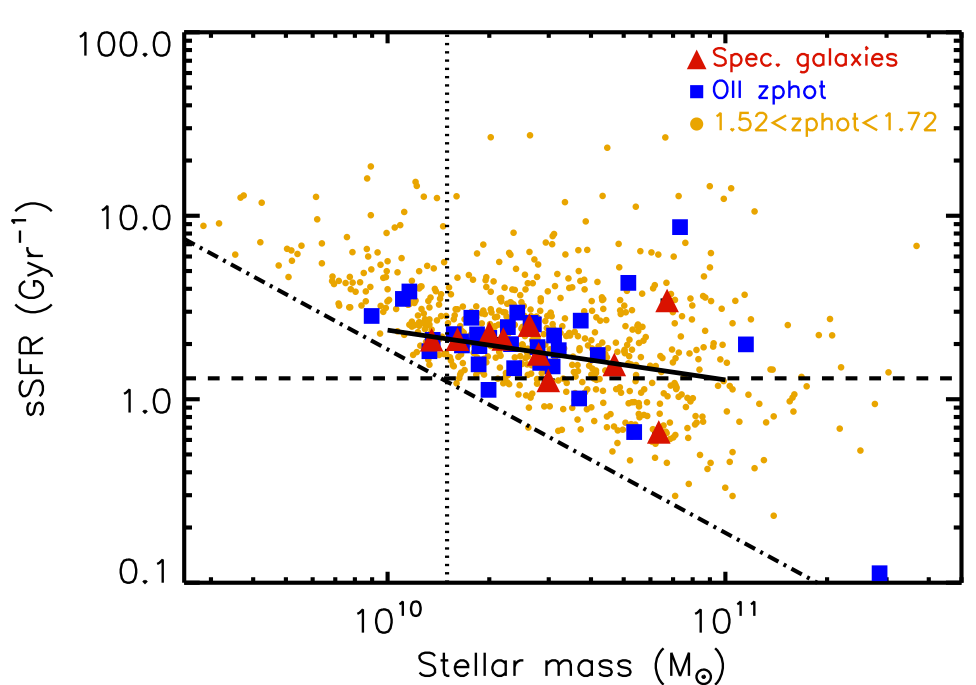} 
\includegraphics[width=8.5cm,angle=0]{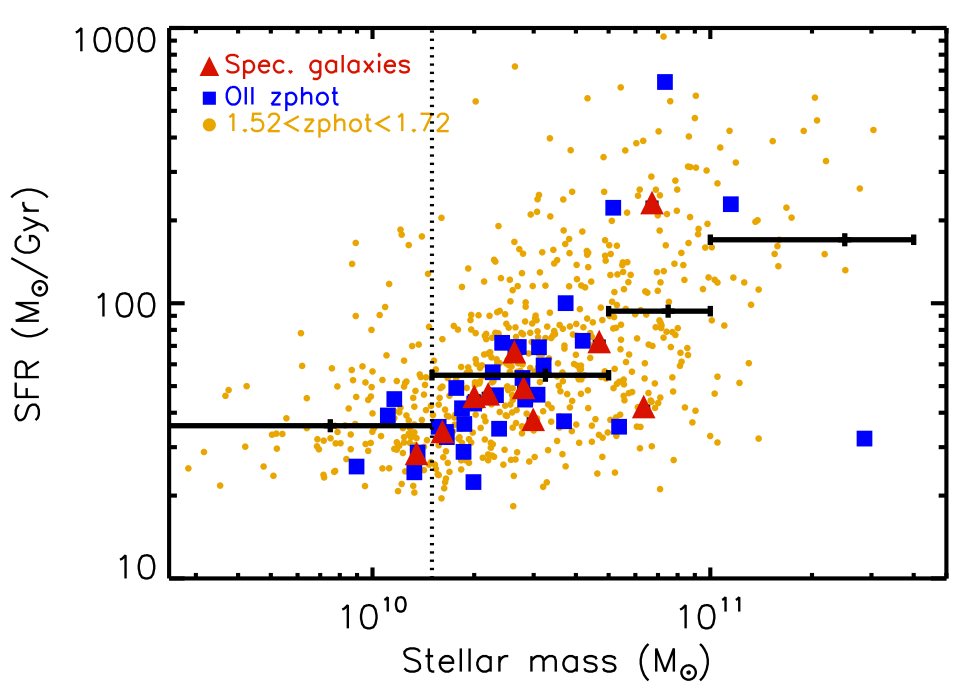} 
 \caption{\textit{Left} Specific star formation rate as function of stellar mass for the three samples of galaxies 1) spectroscopically galaxies at $z\sim$1.62 (red), 
 2) [OII] $z_{phot}$ members (blue) and MIPS emitters with 1.52$<z<$1.72 (orange). The vertical dotted line indicates 
 our mass completeness limit; the horizontal dash line marks the main-sequence level and the diagonal dash-dot line establishes the 
 5-$\sigma$ sensitivity limit of our MIPS data. The solid black line is a fit to the 14 member galaxies within the cluster region ($r<$ 1 Mpc).
 \textit{Right} Star formation rate as function of stellar mass. The median SFR of the full sample per mass bin in indicated by the horizontal black solid lines.  
 }
 \label{ssfr}
\end{figure*}

We first explore the relation between the stellar mass and SFR. 
In Fig. \ref{ssfr} we plot the specific star formation rate (sSFR=SFR/$M_{*}$) versus the stellar mass for the three galaxy samples used in our study.
The deep 24$\mu m$ data allows us to reach the main sequence in this distant cluster (horizontal dash line in the figure), 
sSFR$_{MS}$= 1.3 Gyr$^{-1}$ obtained with the relation proposed by \cite{Elbaz11}, $sSFR_{MS}\,\, [Gyr^{-1}] = 26 \times t_{cosmic}^{-2.2} $. 
The majority of the (U)LIRGs in the CLG0218 system and environment are therefore starbursts (see Fig. \ref{ssfr}). 
As shown in Table 3 and Fig. \ref{ssfr}-right panel, we find that the median SFR increases with increasing stellar mass. 
This trend as well as the absolute median SFR values are very similar for the 3 galaxy samples. 
Given the low number statistics of the spectroscopic and [OII] emitter samples, we will use from now on the overall sample of 
693 galaxies detected by MIPS.

The sSFR is essentially a measure of the star formation efficiency of galaxies, 
and the large scatter in the sSFR -- $M_{*}$ relation is generally attributed to complex processes related with the amount of gas available or used 
to convert to stars (e.g. Salmi et al. 2012). A negative trend in the sSFR--$M_{*}$ relation has been reported at all redshifts \cite[e.g.][]{Rodighiero}. 
In CLG0218 we confirm this trend, where the most massive galaxies have the lowest sSFR, that implies that more massive galaxies formed their stars 
earlier and more rapidly than their low mass counterparts.

The locus of the 14 star forming cluster galaxies (i.e., within 1 Mpc) forms a tight relation, described by a linear fit with slope $\alpha$ =-0.012,  
 a mean of 2.0 Gyr$^{-1}$, and a standard deviation of 0.89 (see diagonal black solid line in Fig. \ref{ssfr}).   
We compare our results on the sSFR -- $M_{*}$ relation of CLG0218 with other published work.  
The redshift evolution of the specific SFR has been studied in a number of recent papers of which we highlight a recent compilation of data by 
\cite{Sargent}. The authors find an evolutionary trend described by a power law $\propto$ (1+z)$^{2.8\pm0.1}$. At redshift $\sim$1.6 Fig. 1.c of 
\cite{Sargent} predicts a sSFR for galaxies with $M_{*}\sim$5$\times$10$^{10}$M$\odot$ of $\sim$1.34 Gyr$^{-1}$. 
In CLG0218 we measure, for the same stellar mass, and using the linear fit to the 14 cluster members, a sSFR of 1.89 Gyr$^{-1}$, 
a value that is slightly higher than the one expected with the power law fitting function, but closer to the observational results at the same 
redshift from \cite{Elbaz11}. Therefore, the sSFR--$M_{*}$ relationship of the cluster CLG0218 agrees well with the relationship for the GOODS field 
galaxies presented in \cite{Elbaz11}.

\begin{table}
\caption{Star formation rates measured from the MIPS 24$\mu m$ fluxes as function of stellar mass for the 3 samples: 
10 spectroscopic galaxies, [OII] emitters, photometric redshift candidates with 1.52 $< z_{phot}<$ 1.7.}  
\label{table:3}     
\small
\centering           
\begin{tabular}{lll} 
\hline\hline   
Sample           & Mass bin         &        median SFR   \\
                       & (M$_\odot$)    &        (M$_\odot$/yr)   \\
\hline                         
spec.        &  1.5--5$\times$10$^{10}$  &      47          \\
spec.        &  5--10$\times$10$^{10}$   &    191            \\
spec.        &   $\ge$10$^{11}$               &    --            \\
\hline
OII, $z_{phot}$              &   1.5--5$\times$10$^{10}$  &    46            \\
OII, $z_{phot}$               &   5--10$\times$10$^{10}$   &    222            \\
OII, $z_{phot}$               &   $\ge$10$^{11}$                &    229            \\
\hline
$z_{phot}$         &  1.5--5$\times$10$^{10}$   &     56           \\
$z_{phot}$        &  5--10$\times$10$^{10}$    &     92          \\
$z_{phot}$         &   $\ge$10$^{11}$                &     161          \\
\hline
\end{tabular}
\end{table}

\section{Mapping the SFR in the cluster and its environment}

In this section we investigate the relation between SF, environment and galaxy stellar mass. 
We explore two main approaches to define and study the environment: i) fixed aperture, ii) local galaxy density. We also study the spatial 
distributions of the star forming and passive galaxies at 1.52$<z_{phot}<$1.72. Finally, we calculate the normalized total cluster SFR and compare 
it with works in the literature to gain insight on the redshift evolution of the SFR in groups and clusters.

\subsection{Star formation rate vs cluster centric distance}

\begin{figure}
\hspace{-0.3cm}\includegraphics[width=9.2cm,angle=0]{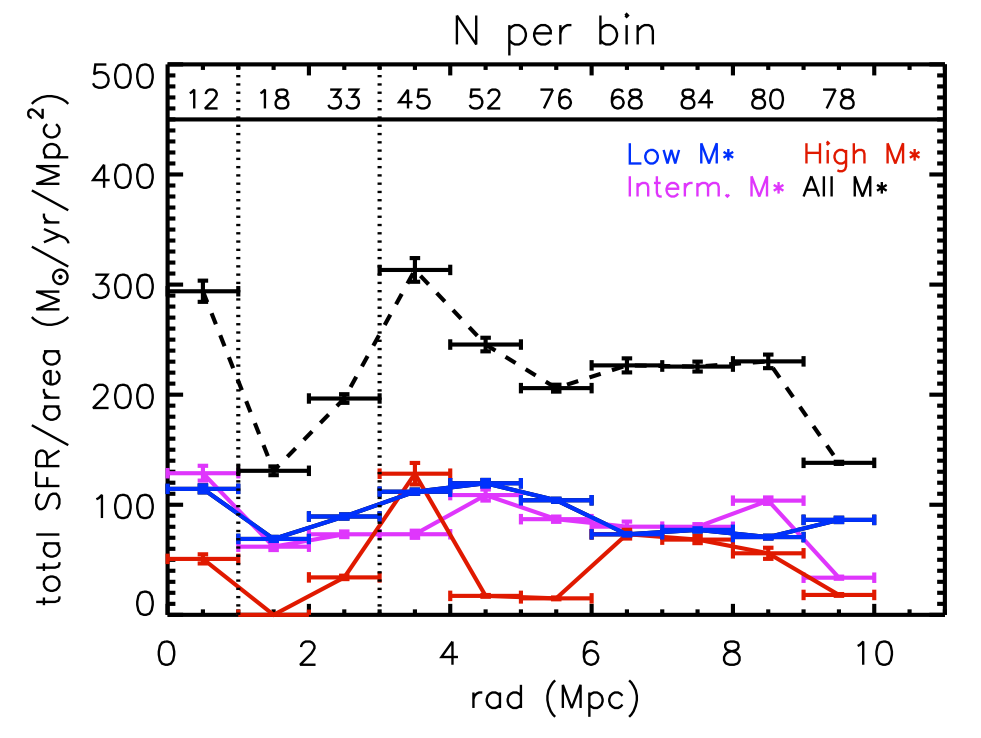} 
 \caption{SFR measured in annuli of 1 Mpc width normalized by the bin area vs cluster centric distance. The SF sample 
 is divided in 3 subsamples according to their stellar mass. The black dash line shows the results for the full (all $M_{*}$) sample, 
 and the corresponding errors take into account the number of galaxies per bin, which are listed at the top of the plot. }
 \label{sfrdist}
\end{figure}

We divide our full sample of 693 star forming galaxies in three mass bins, taking into account our mass completeness level 
of 1.5$\times$10$^{10}$ M$_\odot$: 
1) the low-mass bin, 1.5--5$\times$10$^{10}$ M$_\odot$, 2) an intermediate mass category, 5--10$\times$10$^{10}$ M$_\odot$, 
and 3) massive galaxies with $\ge$10$^{11}$ M$_\odot$.
We then group the galaxies in 1 Mpc slices. As in Tran et al (2010) and Bassett et al (2013), we consider the cluster region 
to be within a radius of 1 Mpc. As suggested by Bassett et al. and Tadaki et al, the field is defined by the galaxies beyond 
a radius of 3 Mpc cluster centric distance. 
Figure \ref{sfrdist} shows the total SFR in each bin, normalized by the bin area for the 
three stellar masses regimes and the combined sample. The low and intermediate mass curves follow similar trends and absolute
values, with a sharp transition at a radius 1--2 Mpc where the SFR decreases by a factor $\sim$2 relative to the inner bin, and 
then increases at $r$ = 4--5 Mpc, stabilizing to a level of 70--90 M$_\odot$/yr/Mpc$^{2}$.
The high stellar mass galaxies present a different behavior, though we are limited by small number statistics. The inner bin with 4 galaxies 
has a lower value of 50 M$_\odot$/yr/Mpc$^{2}$, immediately 
followed by  a sharp decline with no galaxies in the 1--2 Mpc bin. From then on we see a marked increase in the total SFR reaching 120 M$_\odot$/yr/Mpc$^{2}$ 
at $r$= 3--4 Mpc, again a decrease to nearly 0, resuming to a total SFR per area at r$>$6 Mpc  concordant with the lower stellar mass bins.

This figure suggests that galaxy stellar mass may play an important role in the distribution / environment of star-forming galaxies in clusters.
In particular, high $M_{*}$ star forming galaxies have a low SFR contribution in the cluster region and are nearly absent in the in-falling region. 
In addition, we highlight the quenching action of the transition region at 1--3 Mpc, characterized by a lower SFR in star-forming galaxies of any mass bin.

The behavior of the combined stellar mass samples (black dash line in Figure \ref{sfrdist}) mirrors the trends described above. 
The cluster bin has a high SFR value of $\sim$300 M$_\odot$/yr/Mpc$^{2}$ which plunges down to 130 M$_\odot$/yr/Mpc$^{2}$ at 1--3 Mpc 
but rises to approximately the cluster value at $r$= 4 Mpc and then stabilizes to an intermediate value of $\sim$250 M$_\odot$/yr/Mpc$^{2}$. 
The main caveat to the interpretation of this result is that large scale substructure, ie. filaments, may bias this view at large projected distances, 
therefore we investigate in the next section the local galaxy density which is more insensitive to such effects.

\subsection{Star formation rate and galaxy density}
 
One of the most widely used techniques to study the environment in and around galaxy clusters is the use of a local density \citep[e.g.,] [] {Tanaka}
defined as the number of galaxies within a circle of radius equal to the distance of the $n$th neighbour, normalized by the enclosed area. To be
consistent with previous work we consider a local density defined with $n$=5, $\Sigma_{5}$. To compute the local density we use the combined sample 
of star forming and passive galaxies that lie above the mass cut $M_{*}$$>$1.5$\times$10$^{10}$M$_\odot$. 
 
\begin{figure*}
\includegraphics[width=14cm,angle=0]{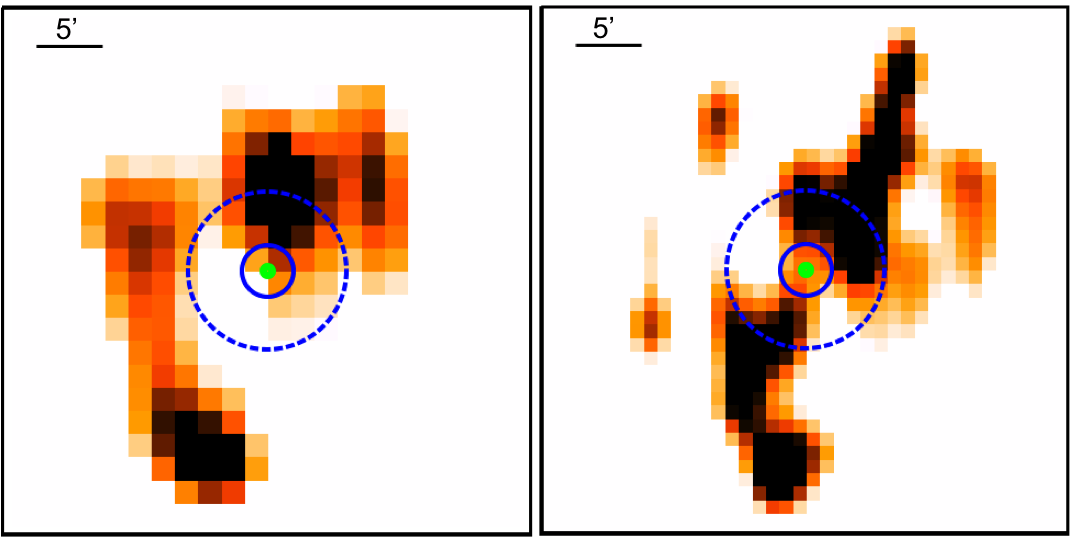} 
 \caption{ Gaussian smoothed density maps of the (left) passive galaxies and (right) star forming galaxies obtained with the distance to the 5th 
 nearest neighbour. Image sizes are 40$\arcmin \times$40$\arcmin$. North is up and East is to the left. The images are centered on ID 16 
 (green circle), the two blue circles represent radii of 3 Mpc (dash circle) and 1 Mpc (solid). Two significant filamentary 
 structures are seen in the upper-right and lower-left quadrants in both the passive and SF galaxy maps.}
\label{densitymap}
\end{figure*}

We find no obvious trend between SFR and local density, in line with the result found by \cite{Tadaki} using a smaller and 
somewhat different sample including the [OII] emitters, and more recently \cite{Ziparo}.
However, the galaxy density maps computed with 
the distance to the 5th nearest neighbour for the star forming and passive galaxies independently, show two large filamentary structures several 
Mpc across (Fig. \ref{densitymap}). This confirms the view that groups and clusters are found at the crossings of such matter filaments and what is 
generally termed the field is not simply a random distribution of galaxies.

 \begin{figure*}
\includegraphics[width=9.2cm,angle=0]{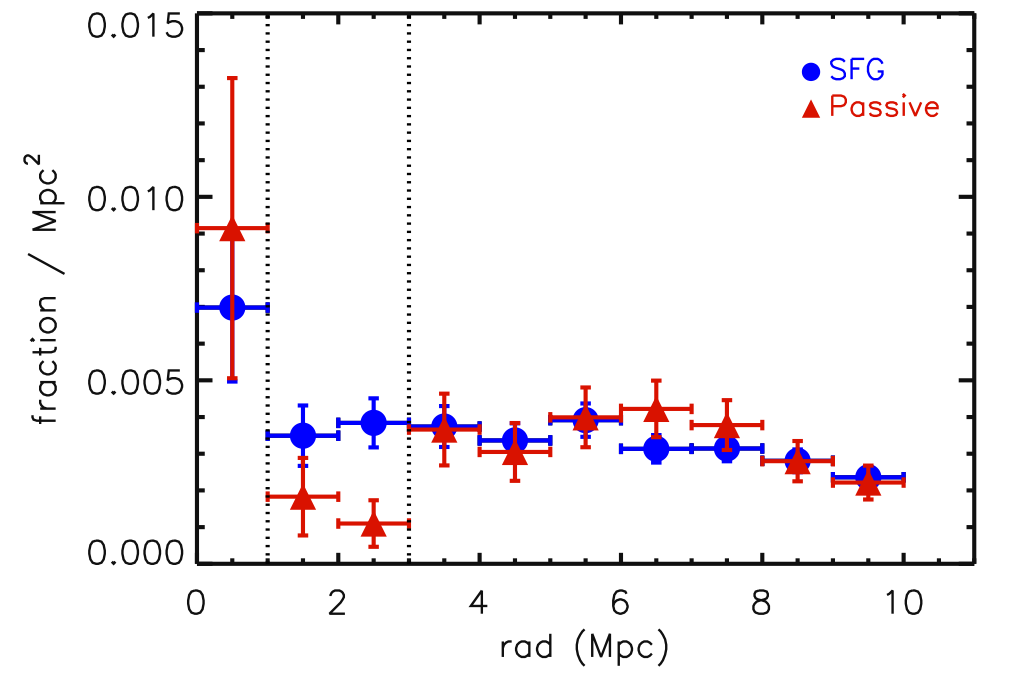} 
\includegraphics[width=8.4cm,angle=0]{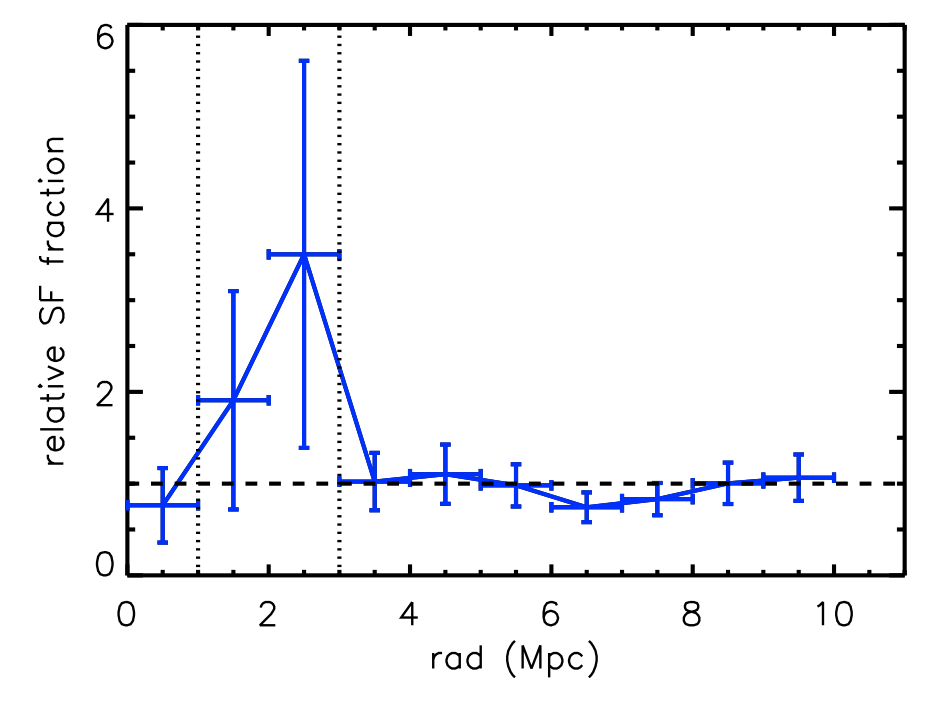}
 \caption{(Left) Fraction of star forming (blue) and quiescent (red) galaxies measured in 1 Mpc slices versus projected cluster 
 concentric distance (Right) Ratio of the two curves displayed on the left panel. }
 \label{fractionpassive}
\end{figure*}
 
 \subsection{The relative fraction of SF to passive galaxies}
 
 Besides studying the star-forming population of this cluster and its surrounding region, we also investigate the relative fraction of SF to 
 passive galaxies in the same area. This is important because in the lower redshift Universe galaxy clusters are typically dominated 
 by quiescent galaxies, particularly in the core \citep{Dressler, Wilman}. Therefore the study of the fraction of star forming galaxies 
 covering such a large area (i.e., reaching the low galaxy density of the field) enables us to investigate a potential reversal of the SF-density relation 
 at high-redshift.  
 
 The selection of passive galaxies was done using our optical/infrared SED catalogue. We selected galaxies with the same $z_{phot}$ range 
 considered for the star forming galaxies, 1.52--1.72, and we excluded all sources with sSFR $>$0.01 Gyr$^{-1}$.
 Using these criteria we end up with a sample of 205 quiescent galaxies. By applying the mass cut $M_{*}$$>$1.5$\times$10$^{10}$M$_\odot$ we 
 finally have a sample of 174 quiescent galaxies, in contrast with the 547 star forming galaxies above the mass completeness level.
 
 On the left panel of Fig. \ref{fractionpassive} we plot the relative fractions of passive and star forming galaxies (normalizing to the total number of 
 galaxies in each sample) 
 as a function of radius, again by binning the galaxies in 1 Mpc slices and normalized by the area covered by each bin. The errors in each bin are simply 
 the Poisson errors. While in the cluster region ($r<$1 Mpc) and in the field area ($r>$3 Mpc) there is no statistical difference between the two samples, 
 we find an excess of the SF fraction relative to the corresponding passive fraction in the intermediate bin, 1$<r<$3 Mpc. 
 This behavior is quantified in the right-hand panel of Figure \ref{fractionpassive}, where we plot the ratio between the 2 samples. 
 The fraction of star forming galaxies in the transition region is $\sim$4 times larger than that of the passive galaxies, albeit the large associated errors. 
 
Another way to compare the spatial distribution of the star forming and passive galaxies that does not rely on binned data is to compute the two-sided 
Kolmogorov-Smirnov (K-S) statistic of the 2 samples. This approach allows us to avoid any dependency on the bin size and consequently the statistical errors.
We find that the probability of the two distributions being drawn from the same parent sample is 0.24 if we limit the test to a projected distance of 0--3 Mpc, 
and 0.27 when considering the full radial range. Such low probability values strengthen our previous result of significant differences between the samples of passive 
and star forming galaxies. 
 
In addition, we find that the fraction of star forming galaxies is higher in the cluster than anywhere in the field 
(see blue line in Fig. \ref{fractionpassive}, left panel). We quantify this 
behavior by performing a linear fit to the fraction of star forming galaxies over the radial range 1--10 Mpc. We obtain a value $f_{SF > 1 Mpc}$ of 0.0041, that 
represents the area normalized mean fraction of star forming galaxies across the transition and field regions. The central value, 
0.0070$\pm$0.0050, is 1.7 times greater than the mean value, $f_{SF > 1 Mpc}$, however this value is hampered by the statistical errors. 
Hence our conclusion that the fraction of star forming galaxies is significantly enhanced in the cluster relative to the field is a tentative one (1-$\sigma$).

  \subsection{The mass-normalized cluster SFR}

The total SFR of the system ($\Sigma$SFR) is obtained by summing up the SFR of all members above the mass limit,  
enclosed in a circle with 1 Mpc radius: total SFR($<$1 Mpc) = 369$\pm$31 M$_\odot$/yr/Mpc$^{2}$. 
We also calculate the total star formation rate within a radius of 0.5 Mpc to be able to compare our results with \cite{Tran10}, 
who quote a SFR ($r<$0.5 Mpc) equal to 1740 $M_\odot$/yr/Mpc$^{2}$. We find instead a lower 
value of the star formation in the same region: 990$\pm$121 $M_\odot$/yr/Mpc$^{2}$, using \textit{Herschel} SFRs where possible. 
This discrepancy can be understood considering that Tran et al used the \cite{Chary} templates that overestimate $L_{IR}$ at 
these redshifts. 
If we use SFRs from MIPS only we get a lower value, 786 $M_\odot$/yr/Mpc$^{2}$. 
Assuming that the SFRs derived from \textit{Herschel} are the most accurate, this points to an overcorrection of the 
24$\mu m$ star-formation rates when using the Rujopakarn et al method.

The mass-normalized cluster SFR is obtained by dividing $\Sigma$SFR ($<$1 Mpc) by the system gravitational mass, 
5.7$\pm$1.4$\times$10$^{13}$M$_\odot$ \citep{Tanaka10}. We thus obtain, $\Sigma$SFR/M$_{g}$ = 6.5$\times$10$^{-12}$ yr$^{-1}$Mpc$^{-2}$.
The parameter $\Sigma$SFR/M$_{g}$ has been used by several authors to quantify the evolution of the global star-formation rate in clusters 
with redshift. \cite{Koyama11} confirmed previous findings on the rapid increase of the mass-normalized SFR with redshift, 
following the relation $\propto$ (1+$z$)$^{6}$, though his study was based on a small sample of a dozen clusters reaching only $z\sim$ 1 and 
the aperture considered to compute the total SFR was 0.5 Mpc, instead of the virial radius.
More recently \cite{Webb} performed a similar study using the Red Sequence Cluster sample with 42 clusters between 0.3$<z<$1.0 and
found a similar evolutionary trend with a slope of 5.4$\pm$1.9. The authors used $r_{200}$ to compute the total SFR of the cluster and 
attribute the evolution to variations in the in-falling field galaxy population.

For consistency, since we consider the total cluster gravitational mass, we consider as well a radius that encompasses the total cluster mass, 1 Mpc.
Comparing our value, $\Sigma$SFR/M$_{g}$ =6.7$\times$10$^{-12}$ yr$^{-1}$Mpc$^{-2}$, with Figure 11 and 12 of \cite{Koyama11}, 
we find that CLG0218 actually falls in very good agreement with the (1+$z$)$^{6}$ relation. 
 Although this comparison is only qualitative, it lends support to the empirically expected significant increase in the the 
 global star formation of CLG0218, relative to lower redshift clusters. However, our investigation of the radial distribution of the SFR 
 presented in Fig. \ref{sfrdist} as well as the fraction of star-forming galaxies depicted in Fig. \ref{fractionpassive} seems to be inconsistent 
 with the hypothesis raised by \cite{Webb} that the in-falling galaxy population is responsible for a high level of SFR in high-$z$ clusters. 
 Instead, the high level of SFR in the CLG0218 at $z$=1.62 is most likely due to a reversal of the SFR--density relation.

\section{Conclusions}

In this paper we perform a new large scale characterization of the dusty star-forming properties 
in and around CLG0218, a low mass galaxy cluster at $z$=1.6, using deep 24$\mu m$ MIPS and \textit{Herschel} 100--500$\mu m$ imaging 
data covering an area of 20$\times$20 Mpc (comoving scale at the cluster redshift). Here we summarize our main findings:

\begin{enumerate}
  \item using the 24$\mu m$ fluxes we measure luminosity based SFRs in the range 18--2500 M$_\odot$/yr for a sample of 693 
  spectroscopic and $z_{phot}$ galaxies across a radius of 10 Mpc from the cluster center; 
  \item for a subsample of 29 galaxies we have robust \textit{Herschel} detections that allows us to assess the MIPS SFRs: 
  although the scatter is large the overall agreement is good. While contamination from neighbouring galaxies may explain SFRs(Herschel)$>$SFRs(MIPS), it's also possible that the methodology proposed by \cite{Rujopakarn13} may overcorrect the SFR for high luminosity galaxies;
  \item we characterize the brightest FIR cluster galaxy, ID 16, a spectroscopic member with $M_{*}$=7$\times$10$^{10}$M$\odot$, SFR(\textit{Herschel})=256$\pm$70 M$_\odot$/yr 
  and a dust temperature of 34.5$\pm$4.2 K. This galaxy is also taken to be the centre of the cluster. Even though there is X-ray emission associated with this 
  galaxy we do not detect a significant AGN contribution to the FIR SED;
  \item we estimate an extinction of 3 magnitudes in [OII] for the galaxies associated with CLG0218 and its environment;
    \item the SFR contribution from high $M_{*}$ galaxies varies significantly with clustercentric distance and is much lower than that of the lower $M_{*}$ galaxies. 
    High $M_{*}$ are absent in the in-falling region at 1--3 Mpc;
   \item we do not find any obvious trend between the individual galaxy star-formation rates and local galaxy density parametrized by $\Sigma_{5}$;
   \item we measure an enhancement by almost a factor 2 in the fraction of star-forming galaxies in the cluster (defined by an aperture with $r$= 1 Mpc) relative 
   to the field (i.e. $r >$3 Mpc);
    \item we find a significant decrease (by a factor 3.5) in the fraction of passive relative to star-forming galaxies in the in-falling region;
        \item we find two large scale filamentary structures in the galaxy density maps of both the passive and the star-forming samples, showing that the galaxy 
        distribution in the field surrounding CLG0218 is not uniform.  
\end{enumerate}

The enhanced fraction of star-forming galaxies at r$<$1 Mpc seen in Fig. \ref{fractionpassive} shows that SF is nearly two times larger in the \textit{higher}
 density environment of the cluster with respect to the \textit{lower} galaxy density of the surrounding field and in this sense it can be interpreted as a 
 reversal of the SF--density relation.
In addition, the similar relative fraction of SF and passive galaxies within the cluster (also seen at $r>$ 3 Mpc) is surprising and indicates this system 
has a young, active population of galaxies.
Our study shows that the in-falling region (1--3 Mpc) is where most differences are seen in terms of dependance with stellar mass, lower SFR and
 deficiency of quiescent galaxies. This result is in stark contrast to the study of XMMUJ2235.3-2557, a massive galaxy cluster at $z$=1.39, in which
 most of the SFR seen by \textit{Herschel} is at $r_{vir}$ and beyond \citep{Santos}. This suggests that as we move from massive distant clusters to 
 even more distant but less massive clusters/groups, star formation migrates from the field, through the outskirts and onto to the cluster core.
 To better understand this effect, future studies of the star formation in high redshift clusters should include the surrounding cluster environment.

\section*{Acknowledgments}

We thank Nick Seymour for advice on the DecompIR package. 
MT gratefully acknowledges support by KAKENHI No. 23740144. 

PACS has been developed by a consortium of institutes
led by MPE (Germany) and including UVIE (Austria); KUL, CSL,
IMEC (Belgium); CEA, OAMP (France); MPIA (Germany); IFSI, OAP/AOT,
OAA/CAISMI, LENS, SISSA (Italy); IAC (Spain). This development has been
supported by the funding agencies BMVIT (Austria), ESA-PRODEX (Belgium),
CEA/CNES (France), DLR (Germany), ASI (Italy), and CICYT/MCYT (Spain).

SPIRE has been developed by a consortium of institutes
led by Cardiff Univ. (UK) and including Univ. Lethbridge
(Canada); NAOC (China); CEA, LAM (France); IFSI,
Univ. Padua (Italy); IAC (Spain); Stockholm Observatory
(Sweden); Imperial College London, RAL, UCL-MSSL,
UK ATC, Univ. Sussex (UK); Caltech, JPL, NHSC, Univ.
Colorado (USA). This development has been supported by
national funding agencies: CSA (Canada); NAOC (China);
CEA, CNES, CNRS (France); ASI (Italy); MCINN (Spain);
SNSB (Sweden); STFC and UKSA (UK); and NASA (USA).

\bsp

\label{lastpage}

\end{document}